\documentclass[fleqn,usenatbib]{mnras}
\usepackage{newtxtext,newtxmath}
\usepackage[T1]{fontenc}
\DeclareRobustCommand{\VAN}[3]{#2}
\let\VANthebibliography\thebibliography
\def\thebibliography{\DeclareRobustCommand{\VAN}[3]{##3}\VANthebibliography}

\usepackage[dvipdfmx]{graphicx}	
\usepackage{amsmath}	
\usepackage{color}
\usepackage{threeparttable}
\usepackage{array,booktabs}
\usepackage{siunitx}

\newcommand{\msun}{{\rm M_\odot}} 
\newcommand{\rsun}{{\rm R_\odot}} 
\newcommand{\yr}{{\rm yr}} 
\newcommand{\kelvin}{{\rm K}} 

\newcommand{\erad}{E_{\rm r}}
\newcommand{\vvec}{\mbox{\boldmath $v$}}
\newcommand{\avec}{\mbox{\boldmath $a$}}
\newcommand{\rvec}{\mbox{\boldmath $r$}}
\newcommand{\evec}{\mbox{\boldmath $e$}}
\newcommand{\gvec}{\mbox{\boldmath $g$}}
\newcommand{\Pvec}{\mathbf{P}}
\newcommand{\Ivec}{\mathbf{I}}
\newcommand{\Prvec}{\mathbf{P}_{\rm r}}
\newcommand{\sgmvec}{\boldsymbol{\sigma}}
\newcommand{\Fvec}{\mbox{\boldmath $F$}}

\newcommand{\Omgvec}{\mbox{\boldmath $\Omega$}}

\newcommand{\rin}{r_{\rm in}}
\newcommand{\rout}{r_{\rm out}}

\newcommand{\ms}{M_\ast}
\newcommand{\mbh}{M_\bullet}

\newcommand{\mdott}{\dot{M}_{\rm t}}
\newcommand{\medd}{\dot{M}_{\rm Edd}}
\newcommand{\ledd}{L_{\rm Edd}}

\newcommand{\IFR}{\dot{M}_{\rm in}}
\newcommand{\OFR}{\dot{M}_{\rm out}}

\newcommand{\rsph}{R_{\rm sph}}
\newcommand{\rc}{R_{\rm c}}
\newcommand{\rd}{R_{\rm c}}

\newcommand{\red}[1]{{\textcolor{red}{#1}}}

\defcitealias{Wise2019Nature}{W19}
\defcitealias{Sakurai2020MNRAS}{SHI20}
\defcitealias{Toyouchi2020MNRAS}{Toyouchi et al. 2020}

\title[Super-Eddington mass transfer in close binaries]{Radiation hydrodynamical simulations of super-Eddington mass transfer and black hole growth in close binaries}

\author[D.~Toyouchi et al.]{
Daisuke~Toyouchi$^{1,2}$\thanks{E-mail: toyouchi@astro-osaka.jp},
Kenta~Hotokezaka$^{2}$,
Kohei~Inayoshi$^{3}$,
Rolf~Kuiper$^{4}$
\\
$^{1}$Theoretical Astrophysics, Department of Earth \& Space Science, Graduate School of Science, Osaka University, \\
1-1 Machikaneyama, Toyonaka, Osaka, 560-0043, Japan\\
$^{2}$Research Center for the Early Universe (RESCEU), The University of Tokyo,
Hongo, 7-3-1, Bunkyo-ku
Tokyo, 113-0033, Japan\\
$^{3}$Kavli Institute for Astronomy and Astrophysics, Peking University, Beijing 100871, China\\
$^{4}$Faculty of Physics, University of Duisburg-Essen, Lotharstra\ss e 1, D-47057 Duisburg, Germany
}

\date{Accepted XXX. Received YYY; in original form ZZZ}

\pubyear{2023}

\begin{document}
\label{firstpage}
\pagerange{\pageref{firstpage}--\pageref{lastpage}}
\maketitle

\begin{abstract}

Radiation-driven outflows play a crucial role in extracting mass and angular momentum from binary systems undergoing rapid mass transfer at super-Eddington rates.
To study the mass transfer process from a massive donor star to a stellar-mass black hole (BH), we perform multi-dimensional radiation-hydrodynamical simulations that follow accretion flows from the first Lagrange point down to about a hundred times the Schwarzschild radius of the accreting BH.
Our simulations reveal that rapid mass transfer occurring at over a thousand times the Eddington rate leads to significant mass loss from the accretion disk via radiation-driven outflows.
Consequently, the inflow rates at the innermost radius are regulated by two orders of magnitude smaller than the transfer rates.
We find that convective motions within the accretion disk drive outward energy and momentum transport,
enhancing the radiation pressure in the outskirts of the disk and ultimately generating large-scale outflows with sufficient energy to leave the binary.
Furthermore, we observe strong anisotropy in the outflows, which occur preferentially toward both the closest and furthest points from the donor star. 
However, when averaged over all directions, the specific angular momentum of the outflows is nearly comparable to the value predicted in the isotropic emission case.
Based on our simulation results, we propose a formula that quantifies the mass growth rates on BHs and the mass loss rates from binaries due to radiation-driven outflows. 
This formula provides important implications for the binary evolution and the formation of merging binary BHs.

\end{abstract}




\section{Introduction}\label{sec:intro}

Mass transfer from a star to a compact object in binary systems not only results in rich observational phenomena, e.g., X-ray binaries and classical novae, but also plays a crucial role in binary stellar evolution \citep[see, e.g.][]{tauris2023}. 
During mass transfer, the mass and spin angular momentum of each stellar component, as well as the orbital separation, are expected to evolve. 
Therefore, mass transfer is one of the key aspects to understanding the final outcomes of stellar binaries such as merging binary black holes \citep[e.g.][]{Inayoshi2017MNRAS,vandenHeuvel2017MNRAS,van_Son2022bApJ, Briel2023MNRAS}.

The rate of mass supply from a massive star to a compact object can 
be extremely rapid. Depending on the binary mass ratio and the conditions of the massive donor star, such a rapid mass transfer is unstable, leading to a common envelope phase \citep[see, e.g.][]{tauris2023}.
For black hole (BH)-massive star binaries, however, it has been suggested that a good fraction of them avoid unstable mass transfer \citep{Pavlovskii2017MNRAS,vandenHeuvel2017MNRAS}. 
In such systems, the mass transfer rate from a massive stellar companion through the first Lagrange (L1) point often far exceeds the Eddington rate of the accreting BH. 
An example of the systems exhibiting a supper-Eddington mass supply
is an X-ray binary SS 433, which consists of an A-type supergiant donor star and a BH \citep{Hillwig2008ApJ}. The system shows a precessing jet with a velocity of $\sim 0.26c$ and a strong disk wind with a mass loss rate of $\sim 10^{-4}\,{\rm M_{\odot}yr^{-1}}$ with velocities of $100$--$1300\,{\rm km~s^{-1}}$ \citep{Fabrika2004ASPRv}. 
Note that this rate indeed greatly exceeds the Eddington rate of the BH  $\sim 10^{-8}\,{\rm M_{\odot}yr^{-1}}$. In addition, outflows from a supper-Eddington disk have been observed in ultra-luminous X-ray sources \citep[e.g.][]{King2023NewAR}. Therefore, super-Eddington mass transfer is highly non-conservative,
in which outflows and/or jets from the accretion disk remove mass and angular momentum from the binary system. 

The question is, how much mass is accreted onto a BH during stable mass transfer? \cite{Shakura1973A&A} argued that 
the inward mass flux within a super-Eddington disk is proportional to the distance from the central BH as $\dot{m}(r)\propto r$ inside the spherization radius.
In this model, the accretion rate onto the BH is limited to the Eddington rate, and thus, the BH growth is negligibly small over the time scales of massive star evolution.
The hydrodynamic solutions of adiabatic flows have been developed \citep{Ichimaru1977ApJ,ADAF,BB1999,BB04,Quataert2000ApJ} and there has been a long debate about which solution is more realistic \citep{Yuan2014ARA&A}. \cite{BB1999,BB04} showed `adiabatic inflow-outflow solutions', where the outflow is launched from each radius and the disk mass inflow is given by $\dot{M}\propto r^n~(0\leq n < 1)$. \cite{Begelman2012MNRAS} argues that $n=1$ should be achieved for purely adiabatic flows. 
Another class of solutions is a so-called `convection-dominated accretion flow (CDAF)' proposed by \citet{Narayan2000ApJ} and \citet{Quataert2000ApJ}, in which the energy is radially transported due to convection. 
\citet{Inayoshi2018aMNRAS} demonstrated that radiatively inefficient accretion flows from the Bondi radius to a supermassive BH represent the CDAF solution, resulting in $\dot{M}\propto r$ \citep[see also][]{Guo2020ApJ, Xu2023ApJ}{}{}.

Recently, radiation hydrodynamical (RHD) simulations have suggested that the outflow from a super-Eddington disk is weaker than predicted by \citet{Shakura1973A&A}, primarily because the accretion timescale is so short that a significant fraction of photons advect inward without effectively accelerating gas outward \citep{Kitaki2018PASJ,Kitaki2021PASJ,Jiang2019ApJ,Yoshioka2022PASJ}. 
Interestingly, \cite{Yuan2012ApJ} shows with two-dimensional (2D) hydrodynamical simulations for adiabatic flows that the mass inflow index is $n=0.5$--$0.7$. 
The same result is obtained in 2D RHD simulations by \cite{Hu2022ApJ} in the context of supermassive BH accretion. \citet{Kitaki2018PASJ,Kitaki2021PASJ} and \citet{Yoshii2022ApJ} study super-Eddington accretion flows onto a stellar mass black by using 2D RHD. They observe that the outflows from the outer part of the disk failed, leading to a super-Eddington BH accretion.
However, it remains unclear what the stationary configurations of accretion flows with a supper-Eddington mass supply in binary systems would be.


In this paper, we explore stable mass transfer in X-ray binary systems with RHD simulations of the accreting flow including the radiative feedback effects caused by the illuminating inner accretion disk. In order to understand the effect of the Roche potential on the mass outflow, we perform a three-dimensional (3D) simulation that focuses on the outer region of the Roche potential, where non-axisymmetric effects are important. We also conduct a 2D simulation for the inner part to resolve the mass inflow and outflow down to about a hundred times the Schwartzschild radius. 
By combining the 2D and 3D simulations, we aim to unveil the quasi-steady inflow and outflow structures of a super-Eddington accretion disk.

This paper is structured as follows. 
In \S \ref{sec:method}, we describe the numerical method and set-up of our 2D and 3D RHD simulations. 
The main results of our simulations are presented in \S \ref{sec:results}.
Subsequently, in \S \ref{sec:discussion}, we discuss the mass and angular momentum extracted by radiation-driven outflows from mass-transferring binaries.
Finally, our conclusions are summarized in \S \ref{sec:summary}.


\if
Introduce various binary systems with super-Edd. mass transfer rates.
ULX? \cite{Dudik2016ApJ} find an infrared excess from Holmberg IX X-1, which likely arises from a circumbinary disk. \cite{Berghea2020ApJ} obtained an upper limit on the radio flux from a compact radio source in Holmberg IX X-I. This observation rules out the scenario in which the IR excess is powered by a jet. 
S433?
RY Scuti? (Smith, N.)
No theoretical papers?
ULX binary: ultra-luminous X-ray binary.
\fi


\section{Simulation Method}\label{sec:method}

\subsection{Radiation hydrodynamical simulation}\label{sec:RHD}

We employ the hydrodynamical simulation code {\tt PLUTO} 4.1 \citep[][]{Mignone2007ApJS} that we have previously utilized to investigate gas accretion onto isolated BHs \citep[e.g.,][]{Sugimura2017MNRAS, Sugimura2018MNRAS, Toyouchi2019MNRAS, Toyouchi2020MNRAS, Toyouchi2021ApJ, Hu2022ApJ, Inayoshi2022aApJ}. 
In this study, we consider mass transfer in a binary consisting of a BH and a companion star with masses of $\mbh$ and $\ms$, respectively,
and the binary is assumed to be in a circular orbit with a separation of $a$.
Our simulations are conducted in spherical coordinates, ($x$, $y$, $z$) = ($r {\rm sin} \theta {\rm cos} \phi$, $r {\rm sin} \theta {\rm sin} \phi$, $r {\rm cos} \theta$),
co-rotating with the BH around the barycenter of the binary,
such that the BH and the companion star are fixed at $\rvec_\bullet$ = (0, 0, 0) and $\rvec_\ast$ = ($-a$, 0, 0), respectively.

\subsubsection{Basic equations}\label{sec:BEQ}

The basic equations of hydrodynamics we solve are the following:
the equation of continuity,
\begin{eqnarray}
\frac{\partial \rho}{\partial t} + \nabla \cdot (\rho \vvec) = 0,  
\label{eq:mass_cons}
\end{eqnarray}
and the equations of motion,
\begin{eqnarray}
\frac{\partial \rho \vvec}{\partial t} + \nabla \cdot (\rho \vvec \vvec) = \rho \gvec - \nabla \Pvec \ ,  
\label{eq:mom_cons}
\end{eqnarray}
where $\rho$ is the gas density, $\vvec = (v_r, v_\theta, v_\phi)$ is the velocity vector,
$\gvec = (g_r, g_\theta, g_\phi)$ is the sum of the external force vector, $\gvec_{\rm ext}$, and
the radiative force vector exerted on gas by X-ray irradiation from the accreting BH, $\gvec_{\rm irr}$,
$\Pvec = p \Ivec + \Prvec - \sgmvec$ is the stress tensor,
$p$ is the gas pressure,
$\Ivec$ is the identity matrix,
$\Prvec$ is the radiation pressure tensor,
and $\sgmvec$ is the viscous stress tensor.
The equation of state is given with the ideal gas relation,
\begin{eqnarray}
p = (\gamma - 1) \rho \epsilon = \rho \frac{k_{\rm B} T}{\mu_{\rm g} m_{\rm H}} \ ,  
\label{eq:EOS}
\end{eqnarray}
where $\gamma = 5/3$ is the adiabatic index for atomic gas,
$T$ is the gas temperature,
$k_{\rm B}$ is the Boltzmann constant,
$\mu_{\rm g} \sim 0.61$ is the mean molecular weight for fully ionized gas with He-to-H number ratio of 0.1, 
and $m_{\rm H}$ is the mass of hydrogen.

We also solve the equation of total energy,
\begin{eqnarray}
\frac{\partial E}{\partial t} + \nabla \cdot (E \vvec + \vvec \Pvec + \Fvec) = 
\rho \vvec \cdot \gvec + \Gamma_{\rm irr},
\label{eq:ene_cons}
\end{eqnarray}
and the equation of radiation energy,
\begin{eqnarray}
\frac{\partial \erad}{\partial t} = 
- \nabla \cdot (\erad \vvec) - \nabla \vvec \colon \Prvec 
- \nabla \cdot \Fvec + \kappa_{\rm P} \rho c \left ( a_{\rm r} T^4 - \erad \right),
\label{eq:rad_cons}
\end{eqnarray}
where $E = \rho \epsilon + \erad + 1/2 \rho v^2$ is the total energy density, 
$\epsilon$ is the specific internal energy,
$\erad$ is the radiation energy density,
$c$ is the speed of light,
$a_{\rm r}$ is the radiation constant,
$\Gamma_{\rm irr}$ is the net heating rate associated with X-ray irradiation,
$\Fvec$ is the flux of diffusive radiation, and 
$\kappa_{\rm P}$ is the Planck mean gas opacity.
Note here that the first and second terms in the right-hand side of Eq.~(\ref{eq:rad_cons}), which represent the relativistic effects in $O(v/c)$, are not included in the original {\tt PLUTO}.
Therefore, we have incorporated these terms into the implicit radiation transport solver, developed by \citet{Kolb2013A&A}.
In this solver, the equation is discretized at timestep $n$ by regarding $\erad$ on the right-hand side as the value at timestep $n+1$.

\subsubsection{External forces}\label{sec:EF}

In our simulations, the external force vector is given by
\begin{eqnarray}
\gvec_{\rm ext} = - \nabla \Phi_{\rm tot} + \avec_{\rm cf} + \avec_{\rm cr} \ .
\label{eq:gext}
\end{eqnarray}
The first term on the right-hand side represents the gravity of the BH and the companion star,
and $\Phi_{\rm tot}$ is given by
\begin{eqnarray}
\Phi_{\rm tot} = - \frac{G \mbh}{r} - \frac{G \ms}{|\rvec -\rvec_\ast|} \ .
\label{eq:Phi_tot}
\end{eqnarray}
The remaining terms on the right-hand side are the centrifugal force and the Coriolis force emerging from the frame rotation:
\begin{eqnarray}
\avec_{\rm cf} = \Omgvec \times \left ( \Omgvec \times (\rvec - \rvec_{\rm b}) \right ) \ ,
\label{eq:a_cf}
\end{eqnarray}
\begin{eqnarray}
\avec_{\rm cr} = - 2 \Omgvec \times \vvec \ ,
\label{eq:a_cr}
\end{eqnarray}
where $\Omgvec = \sqrt{G M_{\rm tot} / a^3} \evec_{z}$ is the angular velocity of the co-rotating frame, the total mass of the binary is $M_{\rm tot} = \mbh+M_\ast$, and $\rvec_{\rm b} = x_{\rm b} \evec_{x} = -(1-\mu) a \evec_{x}$ represents the location of the barycenter described with $\mu \equiv q/(1+q)$ and $q \equiv \mbh / \ms$.

\subsubsection{Explicit Viscosity}\label{sec:EV}

In our RHD simulations without magneto-hydrodynamical effects, explicit viscosity is imposed to induce angular momentum transport within the accretion disk.
The viscous stress tensor is described by
\begin{eqnarray}
\sigma_{ij} = \rho \Tilde{\nu} \left [ 
\left ( \frac{\partial v_j}{\partial x_i} + \frac{\partial v_i}{\partial x_j} \right ) 
- \frac{2}{3} \left ( \nabla \cdot \vvec\right ) \delta_{ij} 
\right] \ ,
\label{eq:sgm1}
\end{eqnarray}
where $\Tilde{\nu}$ is the shear viscosity, and the bulk viscosity is neglected.
To mimic angular momentum transport by the magneto-rotational instability, non-zero azimuthal components of the shear tensor are assumed as,
\begin{eqnarray}
\sigma_{r \phi} = \rho \Tilde{\nu} r \frac{\partial}{\partial r} \left( \frac{v_\phi}{r} \right) \ ,
\label{eq:sgm2}
\end{eqnarray}
\begin{eqnarray}
\sigma_{\theta \phi} = \rho \Tilde{\nu} 
\frac{{\rm sin}\theta}{r}
\frac{\partial}{\partial \theta} \left( \frac{v_\phi}{{\rm sin}\theta} \right) \ 
\label{eq:sgm3}
\end{eqnarray}
\citep[e.g.,][]{Stone1999MNRAS}.
We present the strength of anomalous shear viscosity with the $\alpha$-prescription \citep[][]{Shakura1973A&A},
\begin{eqnarray}
\Tilde{\nu} = \alpha \frac{c_{\rm eff}^2}{\Omega_{\rm K}}~{\rm exp} \left ( - \frac{|z|}{H} \right ) \ ,
\label{eq:nu}
\end{eqnarray}
where $\alpha$ is the viscous parameter, $\Omega_{\rm K} \equiv \sqrt{G \mbh / r^3}$ is the Keplerian angular frequency, $c_{\rm eff} \equiv \sqrt{P_{\rm tot}/\rho}$ is the effective sound speed, 
$H \equiv c_{\rm eff}/\Omega_{\rm K}$ is the disk thickness,
and $P_{\rm tot} \equiv \rho k_{\rm B} T / \mu_{\rm g} m_{\rm p} + E_{\rm r}/3$.
The exponential factor ensures that viscosity is active near the equatorial plane, following the prescription in \citet{Inayoshi2022aApJ}.
Throughout this paper, we adopt $\alpha = 0.1$, which is suggested by the observations of ionized outbursting disks in dwarf novae \citep{King2007MNRAS}
and magneto-hydrodynamical simulations that investigated the magneto-rotational instability within accretion disks surrounding BHs \citep[e.g.,][]{Penna2013MNRAS, Hirose2014ApJ}.

\subsubsection{Radiative transfer}\label{sec:RT}

We consider the transfer of diffusive photons emitted by gas, solving Eq.~(\ref{eq:rad_cons}).
We adopt the flux-limited diffusion (FLD) approximation, where the radiative flux is given by 
\begin{eqnarray}
\Fvec = - \lambda \frac{c}{\kappa_{\rm R} \rho} \nabla E_{\rm r} \ ,
\label{eq:Fdiff}
\end{eqnarray}
with the Rosseland mean opacity $\kappa_{\rm R}$.
Here, the flux-limiter $\lambda$ is defined as
\begin{eqnarray}
\lambda = \frac{2+{\mathcal R}}{6+3{\mathcal R}+{\mathcal R}^2} \ ,
\label{eq:lambda}
\end{eqnarray}
using the dimensionless quantity, 
\begin{eqnarray}
{\mathcal R} = |\nabla E_{\rm r}|/(\kappa_{\rm R} \rho E_{\rm r}) \ .
\label{eq:R}
\end{eqnarray}
The radiation pressure tensor is written as
\begin{eqnarray}
{\rm P}_{{\rm r},ij} = {\rm f}_{ij} E_{\rm r} \ ,
\label{eq:Prad}
\end{eqnarray}
with the Eddington tensor,
\begin{eqnarray}
{\rm f}_{ij} = \frac{1}{2}(1-f_0)\delta_{ij} + \frac{1}{2}(3f_0-1) n_i n_j \ ,
\label{eq:f_ij}
\end{eqnarray}
where $f_0 = \lambda + \lambda^2 {\mathcal R}^2$ is the Eddington factor,
and $n_i$ is the $i$-th component of the unit vector in the direction of the radiation energy density gradient.
With Eqs.~(\ref{eq:Fdiff}-\ref{eq:f_ij}), we find ${\rm P}_{{\rm r},ij} \sim (1/3) E_{\rm r} \delta_{ij}$ in the optically thick limit of ${\mathcal R} \rightarrow 0$.
On the other hand, the optically thin limit of ${\mathcal R} \rightarrow \infty$ leads to $|\Fvec| \sim c E_{\rm r}$.
Thus, this formula reasonably reproduces the transition from the optically thick diffusion limit to the optically thin streaming limit.

The opacity we utilize in this study is modeled with the OPAL tables at $T \geq 10^4~\kelvin$ \citep[][]{Iglesias1996ApJ}
and the tables by \citet{Ferguson2005ApJ} at $10^3~\kelvin \leq T \leq 10^4~\kelvin$
\footnote{Both the opacity tables are conveniently compiled by the {\tt MESA} code in the {\tt /mesa/kap/} directory \citep[][]{Paxton2019ApJS}.}.
In our simulations, we employ two models of gas opacity supposing different chemical compositions: solar-abundance gas with mass fraction $X = 0.7$ (H), $Y = 0.28$ (He), and $Z = 0.02$ (metals), and metal-free gas with $X = 0.7$, $Y = 0.3$, and $Z = 0.0$.
A remarkable difference between the two models is an enhancement in opacity by the bound-bound transitions of iron around $T \sim 2 \times 10^5~\kelvin$.
To highlight the influence of heavy elements on the accretion dynamics, we compare the simulation results among the two gas composition models in \S \ref{sec:metal}.

In addition, 
we consider X-ray irradiation with a radiation temperature of $T_{\rm X} = 10^8~\kelvin$,
supposing the inverse-Compton component produced by the hot coronal gas at the vicinity of the BH.
We inject X-ray radiation from the inner boundary located at $r = \rin$ and calculate its outward transfer.
We describe the injected X-ray luminosity with the inward mass flux measured at $\rin$ based on the fitting formula given by \citep{Watarai2000PASJ},
\begin{eqnarray}
L_{\rm X} =  
\begin{cases}
2~f_{\rm X}~\ledd~\left [ 1 + {\rm ln} \left ( \frac{\dot{M}}{2 \medd} \right ) \right ] & (\dot{M} > 2 \medd) \\
f_{\rm X}~\ledd~\frac{\dot{M}}{\medd} & ({\rm otherwise}) \ \ ,
\end{cases}
\label{eq:MtoL}
\end{eqnarray}
where $\ledd$ and $\medd$ are the Eddington luminosity and mass accretion rate defined as 
\begin{eqnarray}
\ledd = \frac{4 \pi G \mbh c}{\kappa_{\rm T}} = 1.1 \cdot 10^6~L_\odot \left ( \frac{\mbh}{34 \ {\rm M_\odot}} \right ) \ ,
\label{eq:ledd}
\end{eqnarray}
\begin{eqnarray}
\medd = \frac{\ledd}{\epsilon_{\rm r} c^2} = 7.5 \cdot 10^{-7}~M_\odot \ {\rm yr}^{-1} \left ( \frac{\mbh}{34 \ {\rm M_\odot}} \right ) \ ,
\label{eq:medd}
\end{eqnarray}
with the opacity of Thomson scattering for fully ionized hydrogen $\kappa_{\rm T} = 0.4~{\rm cm^2~g^{-1}}$
and the radiative efficiency assumed to be $\epsilon_{\rm r} = 0.1$.
The logarithmic form of $L_X$ in the super-Eddington regime implies that the radiative efficiency declines effectively due to the photon trapping effect.
$f_{\rm X}$ represents the energy fraction of the X-ray radiation to the total radiative energy released by gas accretion, which significantly changes depending on mass accretion rates.
In this paper, we set $f_{\rm X} = 0.1$, supposing high/soft states of X-ray binaries,
where the Eddington ratio is $\dot{M}/\medd \gtrsim 0.1$ \citep[e.g.,][]{Gierlinski1999MNRAS, Done2003MNRAS, Remillard2006ARA&A}.

The X-ray radiative flux at the innermost radius is described by supposing an anisotropic radiation field as,
\begin{eqnarray}
F_{\rm X}(\rin,~\theta) =  \frac{L_{\rm X}}{4\pi \rin^2} f(\theta) \ , \ \ f(\theta) \propto {\rm cos}^2 \theta \ ,
\label{eq:flux_rin}
\end{eqnarray}
where the anisotropic factor is normalized, such that $\int f(\theta) {\rm d}\Omega = 1$.
This anisotropic radiation field is motivated by the RHD simulation of super-Eddington accretion flow by \citep{Ohsuga2005ApJ}, implying that photons preferentially escape perpendicular to the accretion disk plane.
The radiative flux declines outward due to energy loss of X-ray photons via Compton scattering as
\begin{eqnarray}
F_{\rm X}(r,~\theta) =  F_{\rm X}(\rin,~\theta)~{\rm exp} ( - \tau_{\rm e})~ \frac{\rin^2}{r^2} \ ,
\label{eq:flux_any_r}
\end{eqnarray}
where the optical depth $\tau_{\rm e}$ is written as
\begin{eqnarray}
\tau_{\rm e} =  \int^r_{\rin} \kappa_{\rm T} \rho (r', \theta) {\rm d} r' \ .
\label{eq:tau_e}
\end{eqnarray}
Then, the radiative force exerted on gas via electron scattering by X-ray irradiation is given by
\begin{eqnarray}
\gvec_{\rm irr} = \frac{\kappa_{\rm T} F_{\rm X}}{c}~\evec_{r} \ .
\label{eq:gvec_irr}
\end{eqnarray}
Note here that our simulations do not include line forces associated with bound-bound absorption by neutral or partially ionized heavy elements,
which could contribute to generating fast outflows \citep[e.g.,][]{Proga2000ApJ, Nomura2013PASJ, Nomura2016PASJ}.
This is justified by \citet{Nomura2021MNRAS}, who shows that mass loss rates by line-force-driven outflows decrease for less massive BHs as $\dot{M}_{\rm out} \propto \mbh^{4/3}$, such that its effect is negligible as long as considering gas accretion onto stellar mass BHs.

Additionally, we describe the net heating rates associated with X-ray irradiation as
\begin{eqnarray}
\Gamma_{\rm irr} = \Gamma_{\rm CP} + \Gamma_{\rm PE} \ ,
\label{eq:gamma_irr}
\end{eqnarray}
where the first term on the right-hand side represents
the sum of Compton heating and inverse-Compton cooling rates, and 
the second one represents the sum of X-ray photoionization heating and recombination cooling rates.
These rates are written with the fitting formula by \citet{Blondin1994ApJ},
\begin{eqnarray}
\Gamma_{\rm CP} = 8.9 \cdot 10^{-36}~n_{\rm H}^2~\xi~(T_{\rm X} - 4 T) \ \ {\rm erg~s^{-1}~cm^{-3}} \ ,
\label{eq:gamma_cp}
\end{eqnarray}
\begin{eqnarray}
\Gamma_{\rm PE} = 1.5 \cdot 10^{-21}~n_{\rm H}^2~\xi^{1/4}~T^{-1/2}~
\left ( 1 - \frac{T}{T_{\rm X}} \right ) \ \ {\rm erg~s^{-1}~cm^{-3}} \ ,
\label{eq:gamma_pe}
\end{eqnarray}
where the ionization parameter is defined as $\xi \equiv 4 \pi F_X /n_{\rm H}$, and
$n_{\rm H}$ is the number density of hydrogen.
Note that while these fitting formulas are originally derived for optically thin gas illuminated by Bremsstrahlung radiation with $T_{\rm X}$, 
we apply them across both optically thin and optically thick regimes without any corrections.
We adopt this simplification because within the accretion disk, where $\tau_{\rm e} \gg 1$, the ionization parameter is typically small
so that the contribution of $\Gamma_{\rm CP}$ and $\Gamma_{\rm PE}$ to the thermal evolution of the accretion disk is generally negligible.

\subsection{Numerical setup}\label{sec:IC}

\begin{figure*}
\begin{center}
\includegraphics[width=1.9\columnwidth]{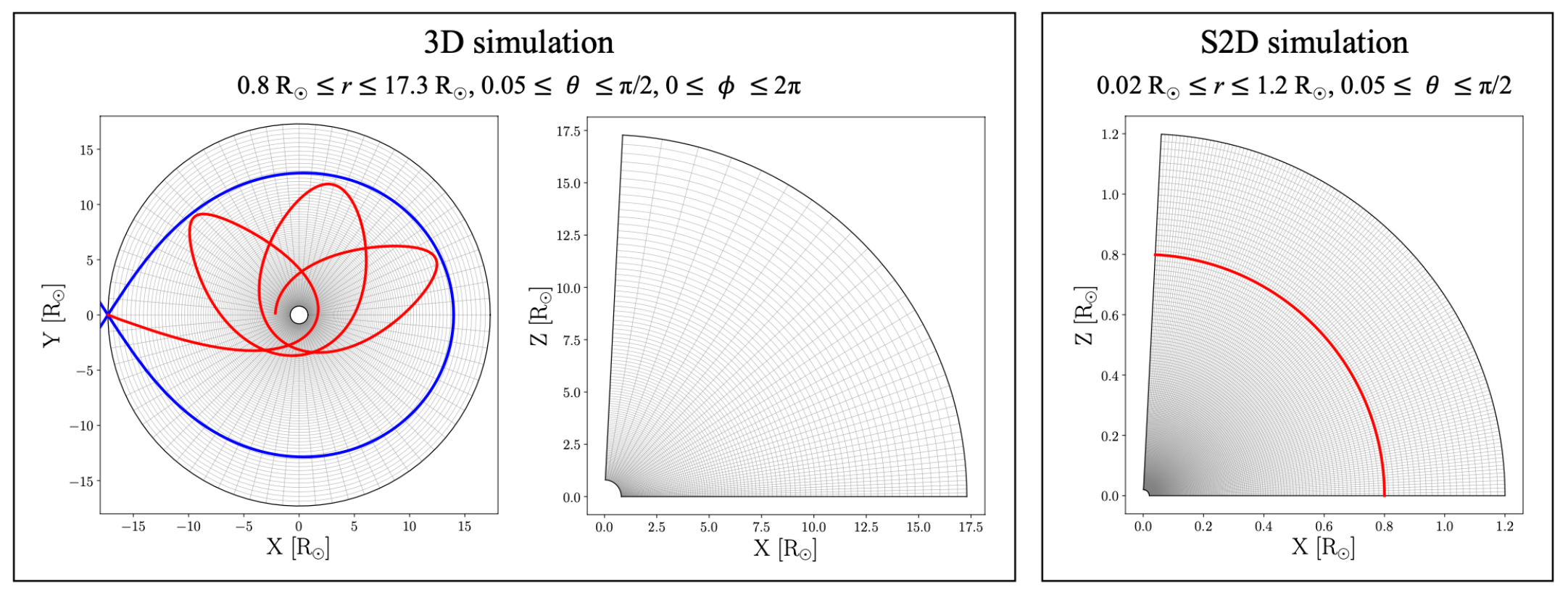}
\end{center}
\vspace{-5mm}
\caption{
Grid configuration adopted in our 3D simulations (left and middle panels) and S2D axisymmetric simulations (right panel).
The black and grey curves represent the boundaries in the numerical domains and those of each cell, respectively.
The blue and red curves in the left panel indicate the Roche lobe containing the L1 point and the ballistic trajectory of gas passing through the L1 point at a radial velocity of $v_r = - 0.01~a \Omega$, respectively. 
The red curve in the right panel represents the effective outer boundary defined as $r^{\rm eff}_{\rm out} = 0.8~\rsun$, which corresponds to the inner boundary of the 3D simulations.
Our analysis for the S2D simulations focuses on $r \leq r^{\rm eff}_{\rm out}$.
Additionally, we conduct the L2D simulations, whose grid configuration is the same as that in the S2D simulations but covers the radial range of $0.2~\rsun \leq r \leq 17.3~\rsun$.
}
\label{fig:grid}
\end{figure*}

Throughout this paper, we choose a BH mass of $\mbh = 34~\msun$, a companion mass of $\ms = 41~\msun$, and an orbital separation of $a = 36~\rsun$.
The binary with these parameters likely undergoes stable mass transfer and subsequently leaves a merging binary BHs, as demonstrated in the Population synthesis model by \citet{Inayoshi2017MNRAS} (see their figures~2 and 3).
When gas transferred from the L1 nozzle approaches the BH, its streamline is bent by the centrifugal force and the Coriolis force in the co-rotating frame.
As a result, the transferred gas effectively acquires the specific angular momentum relative to the BH, $l \approx (1-x_{\rm L1})^2 a^2 \Omega$, 
where $x_{\rm L1}$ is the distance from the donor to the L1 point normalized by the orbital separation, which is approximated with the following polynomial function,
\begin{eqnarray}
x_{\rm L1} \approx -0.355 ({\rm log} q)^2 - 0.251 ({\rm log} q) + 0.5 \ .
\label{eq:x_L1}
\end{eqnarray}
This approximate formula is applicable for $q < 1$, while the L1 position for $q > 1$ is given by $1-x_{\rm L1}(1/q)$
\citep{Frank2002apa..book}.
Then, our binary condition leads to the Keplerian circularization radius of the transferred material, 
\begin{eqnarray}
R_{\rm c} = \frac{l^2}{G \mbh} \approx \frac{(1-x_{\rm L1})^4 a}{\mu} \sim 4.3~\rsun \ ,
\label{eq:r_disk}
\end{eqnarray}
which generally represents the outer edge of the circum-BH accretion disk.
Here, another important scale in our study is the spherization radius $\rsph$, 
where the local radiative flux generated by the disk's viscosity is equal to the local Eddington flux.
In a steady accretion disk with a constant mass flux $\dot{M}$, this radius is written as 
\begin{eqnarray}
\rsph &=&  \frac{3}{4}~\frac{\dot{M} c^2}{L_{\rm Edd}} r_{\rm S} \nonumber \\
&\sim& 1.1~\rsun~\left ( \frac{\dot{M}/\medd}{10^3} \right ) \left ( \frac{\mbh}{34~\msun} \right )^{-1} \ .
\label{eq:R_sph}
\end{eqnarray}
Inside this radius, acceleration by diffusive radiation pressure surpasses the BH gravity, allowing outflows to be launched from the disk surface.
\if
\red{
The spherization radius is comparable to or somewhat larger than the photon trapping radius, approximately given by $R_{\rm tr} \sim (H/R) \rsph$, where $H$ is the disk thickness.
Therefore, radiation-driven outflows start arising from the outer disk regions, where photons are not fully trapped.
}
\fi
In our simulations, we consider two mass transfer rates $\mdott = 10^3~\medd$ and $10^4~\medd$, which represent the two cases of $\rc > \rsph$ and $\rc < \rsph$, respectively.
Note here that $\mdott = 10^4~\medd \sim 0.01~\msun~\yr^{-1}$ roughly corresponds to an ideal upper-limit given by $\mdott \simeq M_\ast/\tau_{\rm KH}$, where $\tau_{\rm KH} \sim 10^3~\yr$ is the Kelvin-Helmholtz time of the companion star.

We investigate the global nature of gas inflows and outflows by performing 3D RHD simulations and two kinds of 2D axisymmetric ones, which cover different radial ranges.
The 3D simulations focus on the outer part of the accretion disk, $(\rin,~\rout) = (0.8~\rsun,~17.3~\rsun)$, to unveil the anisotropic structure induced by the non-axisymmetric external forces in the binary.
In addition, we explore the larger radial range, employing the smaller-scale 2D simulations with $(\rin,~\rout) = (0.02~\rsun,~1.2~\rsun)$
and the larger-scale ones with $(0.2~\rsun,~17.3~\rsun)$, hereafter referred to as the S2D and L2D simulations, respectively.
The S2D simulations focus on the accretion dynamics inside the spherization radius,
whereas the L2D simulations clarify how the outflowing gas generated in the inner disk regions propagates beyond the outer edge of the disk.
In the following, we introduce the grid configuration and the boundary condition adopted in each simulation in detail.

\subsubsection{3D simulations: $0.8~\rsun \leq r \leq 17.3~\rsun$}\label{sec:3D}

The left and middle panels of Figure~\ref{fig:grid} show the numerical domain and grid configuration adopted in the 3D simulation.
We set the innermost radius to $\rin = 0.8~\rsun$, such that the streamline of the transferred gas
does not directly intersect the inner boundary (as shown with the red curve in the left panel).
The outermost radius is consistent with the distance from the L1 point to the BH to treat the Roche lobe overflow as gas supply from the outer boundary without explicitly accounting for the companion star's evolution \citep[cf.,][]{Makita2000MNRAS, Ju2016ApJ, Ju2017ApJ, Pjanka2020ApJ}.
Here, we introduce a nozzle area with an angular extent of ${\rm d}\theta = \pm 0.1$ and ${\rm d}\phi = \pm 0.1$ around the L1 point.
Gas flows through this nozzle area at a radial velocity of $v_{r} = - 0.01 a \Omega$ and zero tangential velocity. 
We normalize the density of the inflowing gas to provide a constant mass transfer rate of $\mdott$
and assume the gas temperature of $T = 10^4~\kelvin$, which roughly corresponds to the effective temperature of the companion star at the onset of mass transfer.

As for the tangential directions, our numerical domain covers $0.05 \leq \theta \leq \pi/2$ and $0 \leq \phi \leq 2~\pi$,
taking an assumption of equatorial symmetry in $\theta$.
We employ the number of grid cells in each direction of $(N_r, N_\theta, N_\phi) = (120, 36, 72)$.
In the $\phi$-direction, we adopt uniformly spaced girds.
In the $r$- and $\theta$-directions, we set up logarithmically spaced grids to ensure a high resolution near the center and equator, allowing at least ten grid cells to resolve the disk thickness.

We adopt outflow boundary conditions at the innermost and outermost cells, where zero gradients across the boundaries are imposed on physical quantities to allow gas to flow out from the computational domain.
On the other hand, gas inflows through the boundaries are prohibited by imposing $v_r \leq 0$ and $v_r \geq 0$ at the inner and outer boundaries, respectively, except for the nozzle area we mentioned above.
For the polar boundary at $\theta = 0.05$, we apply the hollow cone condition introduced by \citet{Sugimura2020bMNRAS}. 
In this condition, we place virtual annular cells at $\theta < 0.05$,
within which the physical values, such as gas density and gas pressure, are calculated self-consistently by solving for gas exchange with the surrounding cells.
This boundary condition allows us to track flows across the pole owing to the non-axisymmetric accretion structure without suffering numerical singularity at $\theta = 0$.
We impose reflective conditions on the equator at $\theta = \pi/2$ and 
periodic boundary conditions for the azimuthal direction at $\phi = 0$ and $2\pi$.

\subsubsection{S2D simulations: $0.02~\rsun \leq r \leq 1.2~\rsun$}\label{sec:S2D}

The numerical domain and grid configuration in the S2D simulations are shown in the right panel of Figure~\ref{fig:grid}.
For the $\theta$-direction, we cover $0.05 \leq \theta \leq \pi/2$, assuming equatorial symmetry, as in the 3D simulations.
As for the $r$-direction, we set the innermost radius to $\rin = 0.02~\rsun \sim 140~r_{\rm S}$, where $r_{\rm S} = 2 G \mbh / c^2$ is the Schwarzschild radius.
We position the outer boundary at $r = 1.2~\rsun$ and consider a continuous gas injection from the area of $\theta \geq \pi/2-0.1$ on the outer boundary with a constant radial velocity of 1~\% of the Keplerian velocity.
The injected gas has a specific angular momentum corresponding to the circularization radius of $1~\rsun$.
Note here that the accretion dynamics outside the circularization radius are considerably affected by the artificial centrifugal barrier.
Therefore, we focus on the accretion dynamics at $r \leq r^{\rm eff}_{\rm out} \equiv 0.8~\rsun$, which corresponds to the innermost radius of the 3D simulations,
and 
the mass injection rates adopted for each model are set such that the inward mass flux at $r = r^{\rm eff}_{\rm out}$ matches that obtained in the 3D simulation with $\lesssim 30~\%$ difference.
Thus, with the 3D and S2D simulations, 
we achieve a seamless depiction of the gas transfer process from the L1 point to about $100~r_{\rm S}$ of the central BH.

For the S2D simulations, we resolve the computational domain in each direction with $(N_r, N_\theta) = (120, 144)$ cells.
In the $r$-direction, logarithmically-spaced grids are employed, 
while uniformly-spaced grids are used in the $\theta$-direction.
The boundary conditions in the S2D simulations remain consistent with those adopted in the 3D simulations, except for the mass injection from the outer boundary.
We note here that the S2D simulation provides substantial outflows from the outer boundary, as shown in \S~\ref{sec:results}.
Strictly speaking, we have to treat such outflows arising from the inner regions as the inner boundary condition in the outer 3D simulations
while this has not been considered in the current numerical setup yet.
We utilize the L2D simulations introduced below to study the potential effect of such gas propagation from the inner to outer disk regions.

\subsubsection{L2D simulations: $0.2~\rsun \leq r \leq 17.3~\rsun$}\label{sec:L2D}

The grid configuration and the boundary condition in the L2D simulations are the same as those in the S2D ones, except for the radial range.
The outermost radius $\rout = 17.3~\rsun$ is identical to that in the 3D simulations.
We set the innermost radius to $\rin = 0.2~\rsun$, about five times smaller than $R_{\rm sph} \sim 1~\rsun$ for $\mdott = 10^3~\medd$.
This allows us to investigate the interaction between outflows arising from $r < \rsph$ and the outer accretion disk.
Here, we consider axisymmetric mass injection from the outer boundary.
The injected gas has a specific angular momentum corresponding to $\rc = 4.3~\rsun$, as given in Eq.~(\ref{eq:r_disk}),
to reproduce the disk formation via Roche lobe overflow.

\renewcommand{\arraystretch}{1.5}
\begin{table*}
\begin{center}
\begin{tabular}[c]{ccccc|cccc} \hline \hline
Model & $\mdott$ & $Z$ & 
$(\rin,~\rout)$ & $(N_r,~N_\theta,~N_\phi)$ & 
$t_{\rm qs}$ & 
$\dot{M}_{\rm in}(\rin)$ & $\dot{M}_{\rm out}(\rout)$ & $l^{\rm avg}_{\rm out}/l_{\rm iso}$ \\
& $[\medd]$ & $[Z_\odot]$ & $[\rsun]$ & & 
$[{\rm day}]$ & 
$[\medd]$ & $[\medd]$ & 
\\ \hline
$\rm M3Z1\_3D$  & $10^3$ & 1 & (0.8, 17.3) & (120, 36, 72) & $40$  & 
$1.1 \cdot 10^3$ & $82$             & $1.33$ \\
$\rm M3Z1\_S2D$ & $10^3$ & 1 & (0.02, 1.2) & (120, 144, 1) & $0.5$ & 
$1.2 \cdot 10^2$ & $1.1 \cdot 10^3$ & $-$ \\
$\rm M3Z1\_L2D$ & $10^3$ & 1 & (0.2, 17.3) & (120, 144, 1) & $30$  & 
$7.2 \cdot 10^2$ & $5.3 \cdot 10^2$ & $-$ \\ \hline
$\rm M4Z1\_3D$  & $10^4$ & 1 & (0.8, 17.3) & (120, 36, 72) & $5$   & 
$7.0 \cdot 10^3$ & $3.8 \cdot 10^3$ & $0.79$ \\
$\rm M4Z1\_S2D$ & $10^4$ & 1 & (0.02, 1.2) & (120, 144, 1) & $0.3$ & 
$5.9 \cdot 10^2$ & $6.9 \cdot 10^3$ & $-$ \\
$\rm M4Z1\_L2D$ & $10^4$ & 1 & (0.2, 17.3) & (120, 144, 1) & $5$   & 
$1.5 \cdot 10^3$ & $8.7 \cdot 10^3$ & $-$ \\ \hline
$\rm M3Z0\_3D$  & $10^3$ & 0 & (0.8, 17.3) & (120, 36, 72) & $80$  & 
$1.1 \cdot 10^3$ & $23$             & $3.21$ \\
$\rm M3Z0\_S2D$ & $10^3$ & 0 & (0.02, 1.2) & (120, 144, 1) & $0.5$ & 
$90$             & $1.4 \cdot 10^3$ & $-$ \\ \hline
$\rm M4Z0\_3D$  & $10^4$ & 0 & (0.8, 17.3) & (120, 36, 72) & $5$   & 
$8.1 \cdot 10^3$ & $2.6 \cdot 10^3$ & $0.30$ \\
$\rm M4Z0\_S2D$ & $10^4$ & 0 & (0.02, 1.2) & (120, 144, 1) & $0.3$ & 
$6.3 \cdot 10^2$ & $7.6 \cdot 10^3$ & $-$ \\
\hline \hline \\
\end{tabular}
\end{center}
\caption{
Models explored in this paper. 
The first column represents the model ID.
From the second to fifth columns, we denote the mass transfer rate, the mass fraction of metals, the radial range, and the number of cells in the numerical domain adopted in each model.
From the sixth to ninth columns, we summarize the results of our simulations.
$t_{\rm qs}$ is the time when the accretion dynamics reaches a quasi-steady state.
$\dot{M}_{\rm in}(\rin)$ and $\dot{M}_{\rm out}(\rout)$ are the inward and outward mass fluxes across the inner and outer boundaries, respectively.
$l^{\rm avg}_{\rm out}$ is the specific angular momentum of outflows defined as Eq.~(\ref{eq:l_iso_tot}).
Note that the quantities listed here are the time-average calculated for $t \geq t_{\rm qs}$.
}
\label{table:models}
\end{table*}
\renewcommand{\arraystretch}{1}

\begin{figure*}
\begin{center}
\includegraphics[width=2\columnwidth]{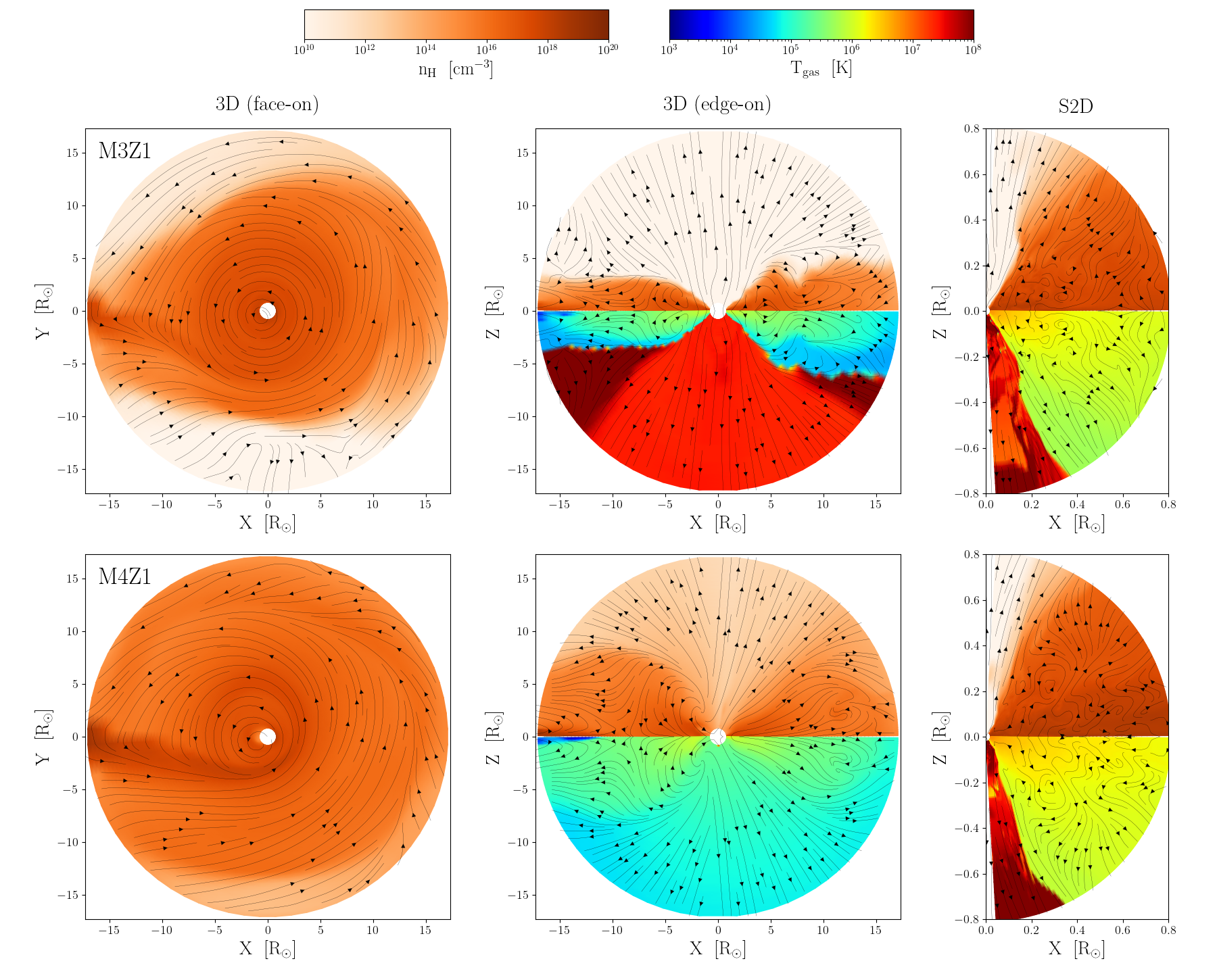}
\end{center}
\caption{
The multi-dimensional structure of gas density and temperature obtained from the M3Z1 model (upper row) and the M4Z1 model (lower row).
The left and middle columns display the face-on and the edge-on views in the 3D simulations, respectively. 
The right column shows the edge-on views in the S2D simulations.
In the face-on maps, only the gas density is shown, while 
the upper and lower halves of the edge-on maps depict the gas density and temperature, respectively.
The black curves with arrows indicate the streamlines.
The snapshots for the M3Z1\_3D (M4Z1\_3D) and the M3Z1\_S2D (M4Z1\_S2D) models are taken at $t = 60~(15)~\rm day$ and $0.8~(0.4)~\rm day$, respectively, representing the quasi-steady state of accretion dynamics.
}
\label{fig:den_temp}
\end{figure*}

\subsection{Models}\label{sec:models}

We explore four basic models with different combinations of two mass transfer rates of $\mdott = 10^3$ and $10^4~\medd$, and two gas compositions of $Z = 0$ and $Z_\odot$.
For each model, we first conduct the 3D simulation until the accretion dynamics reach a quasi-steady state.
Subsequently, we proceed with the S2D simulation, where the mass injection rate is given by the time-averaged inward mass flux measured at the inner boundary of the 3D simulation.
Additionally, we perform the L2D simulation for models with $Z = Z_\odot$, but not for $Z = 0$ to save our numerical resources.

From the first to fifth columns in Table~\ref{table:models}, we give the model ID and numerical setup for each simulation.
In the model ID, the first and second two characters represent the assumed values of $\mdott$ and $Z$, respectively,
and the characters following the underbar represent the grid configuration.
It is worth mentioning here that we also rerun the M4Z1\_3D and M4Z1\_S2D models by doubling the number of grid cells in the tangential directions.
Those simulations with the higher spatial resolution have presented less than 20~\% difference in time-averaged mass fluxes at any radii from the original simulations, suggesting that our simulation results are fairly converged and the conclusions drawn are trustworthy.


\begin{figure}
\begin{center}
\includegraphics[width=\columnwidth]{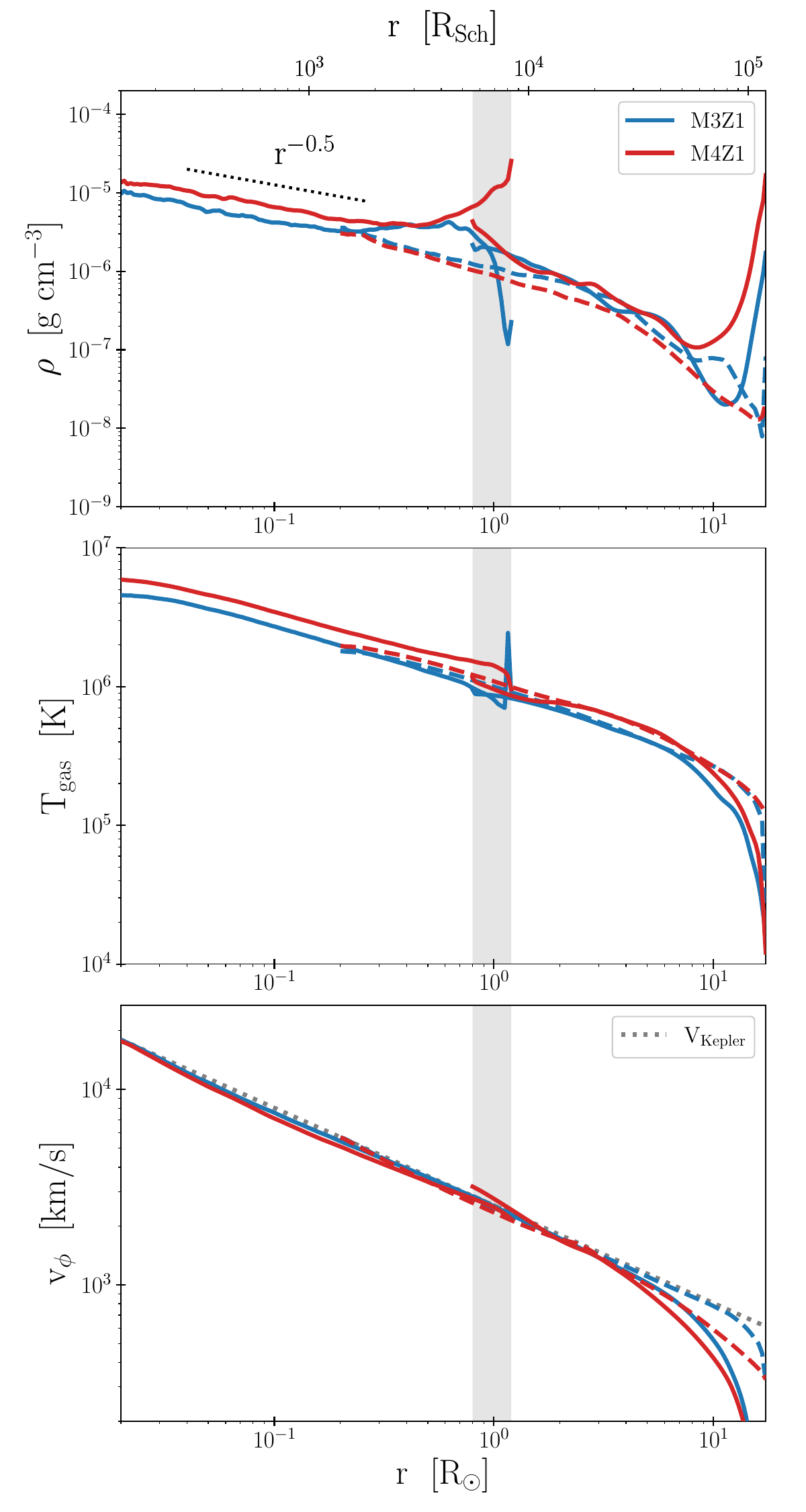}
\end{center}
\caption{
The radial profiles of gas density (top), gas temperature (middle), and rotational velocity (bottom) obtained in our simulations.
The blue and red curves represent the results of the M3Z1 and M4Z1 models, respectively.
The radial profiles obtained in the 3D and S2D simulations are drawn by the solid curves, 
and the radial range where the two simulations overlap, $0.8~\rsun \leq r \leq 1.2~\rsun$, is denoted by the grey-shaded area.
On the other hand, the radial profiles obtained in the L2D simulations are depicted by the dashed curves.
The black-dotted segment in the top panel corresponds to $\rho \propto r^{-0.5}$, the CDAF solution \citep{Narayan2000ApJ, Quataert2000ApJ}.
The grey dotted line in the bottom panel shows the Kepler rotation velocity.
Note here that all profiles are time-averaged for $t \geq t_{\rm qs}$ and also marginalized over the $\phi$-direction for the 3D simulations.
}
\label{fig:rad_profile}
\end{figure}

\begin{figure}
\begin{center}
\includegraphics[width=\columnwidth]{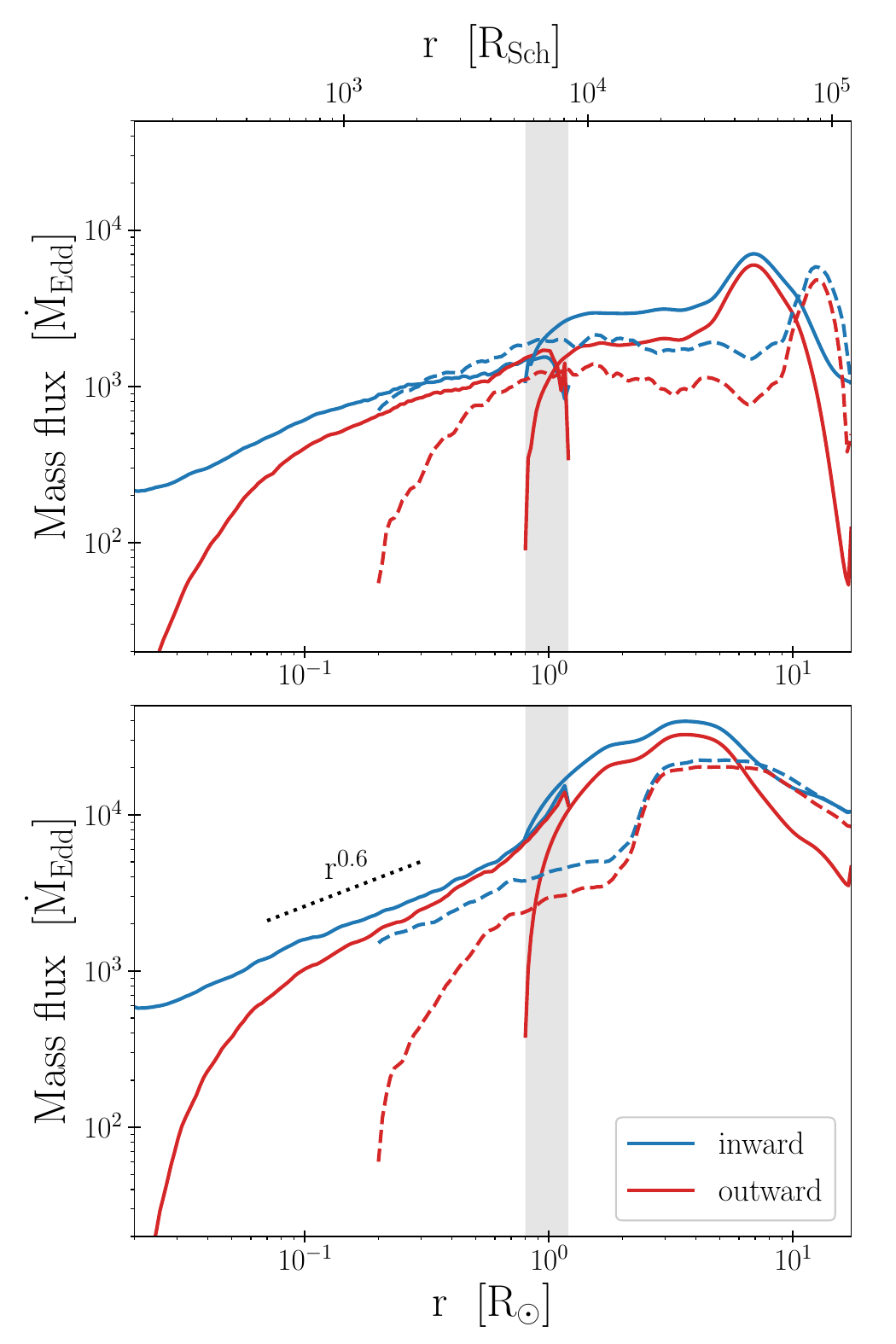}
\end{center}
\caption{
The radial profiles of mass fluxes obtained in the M3Z1 model (upper) and the M4Z1 model (lower), where the mass transfer rates from the L1 point are $\mdott = 10^3~\medd$ and $\mdott = 10^4~\medd$, respectively.
The blue and red curves represent the radial profiles of the inward and outward mass fluxes, respectively.
Then, the solid curves show the results obtained in the 3D and S2D simulations
whereas the dashed ones show those obtained in the L2D simulations.
The black dotted segment denotes a power-law profile of $r^{0.6}$ for comparison with the mass flux profiles in our simulations.
}
\label{fig:mass_flux}
\end{figure}

\begin{figure*}
\begin{center}
\includegraphics[width=1.9\columnwidth]{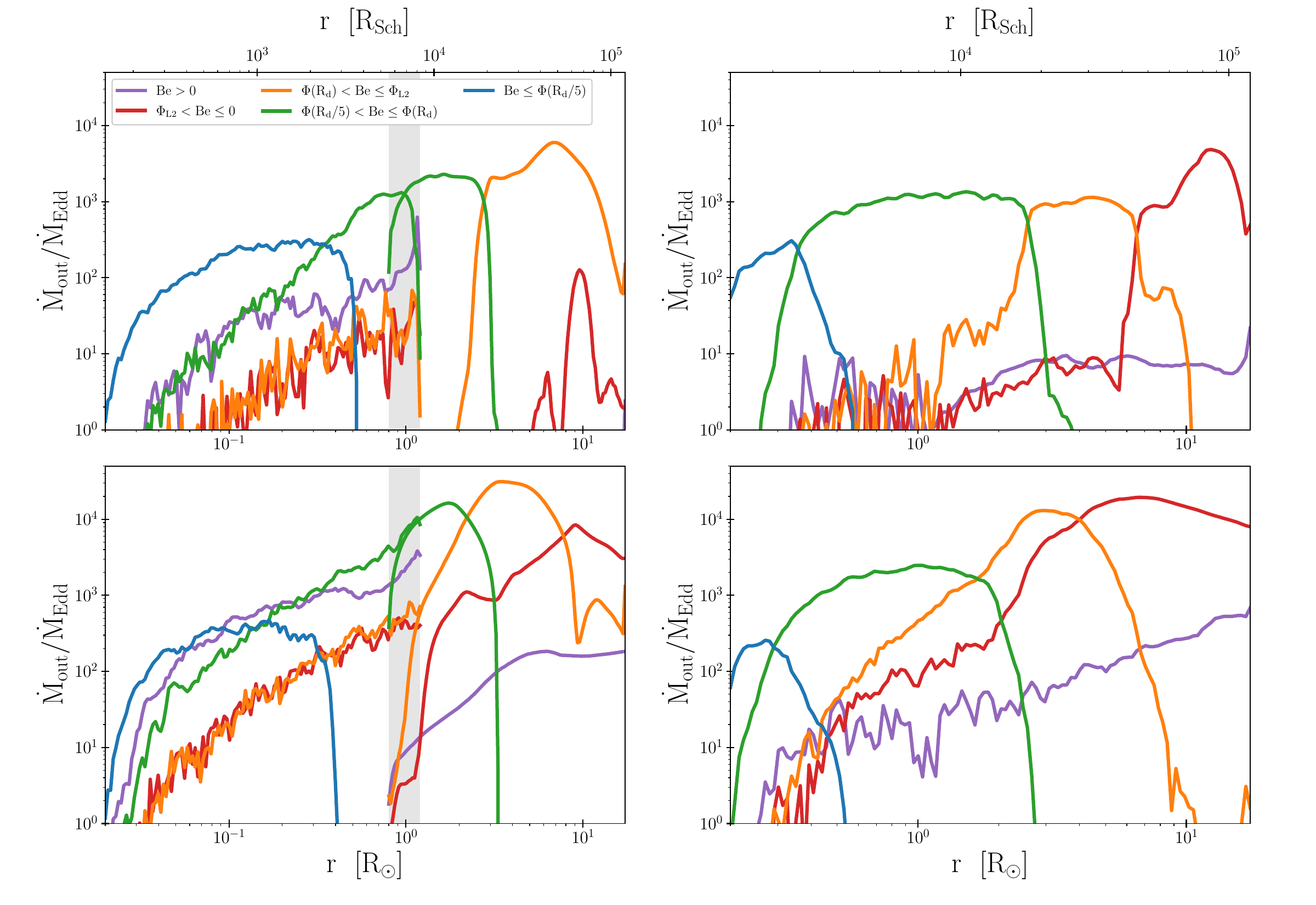}
\end{center}
\caption{
The radial profiles of outward mass fluxes in our simulations separately evaluated for five bins of the Bernoulli numbers; $Be > 0$ (purple), $\Phi_{\rm L2} < Be \leq 0$ (red), $\Phi(\rd) < Be \leq \Phi_{\rm L2}$ (orange), $\Phi(\rd/5) < Be \leq \Phi(\rd)$ (green), and $Be \leq \Phi(\rd/5)$ (blue).
The upper and lower rows correspond to the M3Z1 and M4Z1 models, respectively.
The left column shows the results obtained in the 3D and S2D simulations
whereas the right column shows those obtained in the L2D simulations.
}
\label{fig:mf_Be}
\end{figure*}

\begin{figure*}
\begin{center}
\includegraphics[width=2\columnwidth]{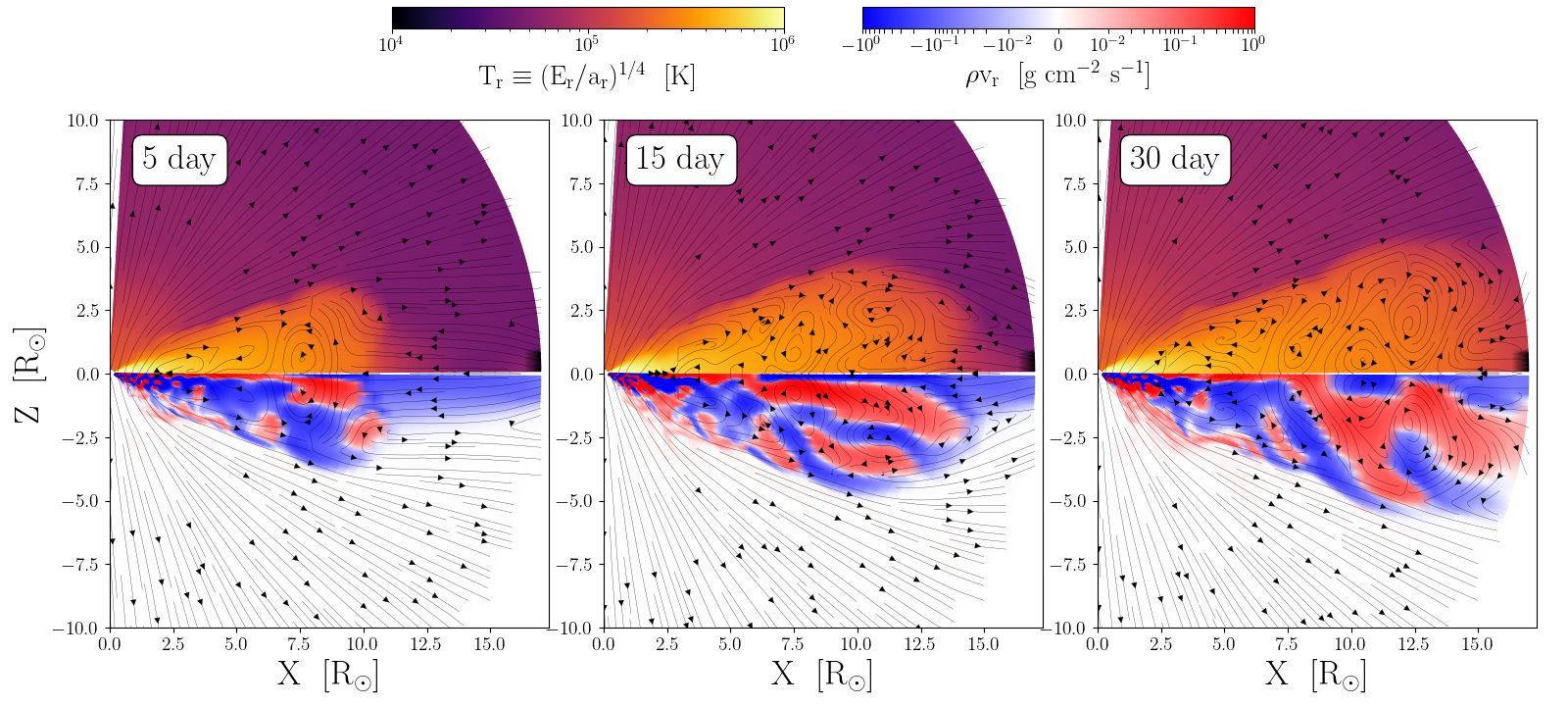}
\end{center}
\caption{
The evolution of hydrodynamic structure obtained in the M3Z1\_L2D simulation.
We present the snapshots taken at $t = 5$, 15, and 30 days from right to left.
The upper halves show the 2D maps of the radiation temperature defined as $T_{\rm r} \equiv (E_{\rm r}/a_{\rm r})^{1/4}$.
The lower halves represent the mass fluxes, with red and blue colors corresponding to outward and inward motions, respectively.
The black curves with arrows indicate the streamlines.
This figure shows that a hot convective region established outside $R_{\rm c} \sim 4.3~\rsun$ gradually extends outward and finally creates large-scale outflows escaping from the outer boundary.
}
\label{fig:den_mf_25D}
\end{figure*}

\begin{figure}
\begin{center}
\includegraphics[width=\columnwidth]{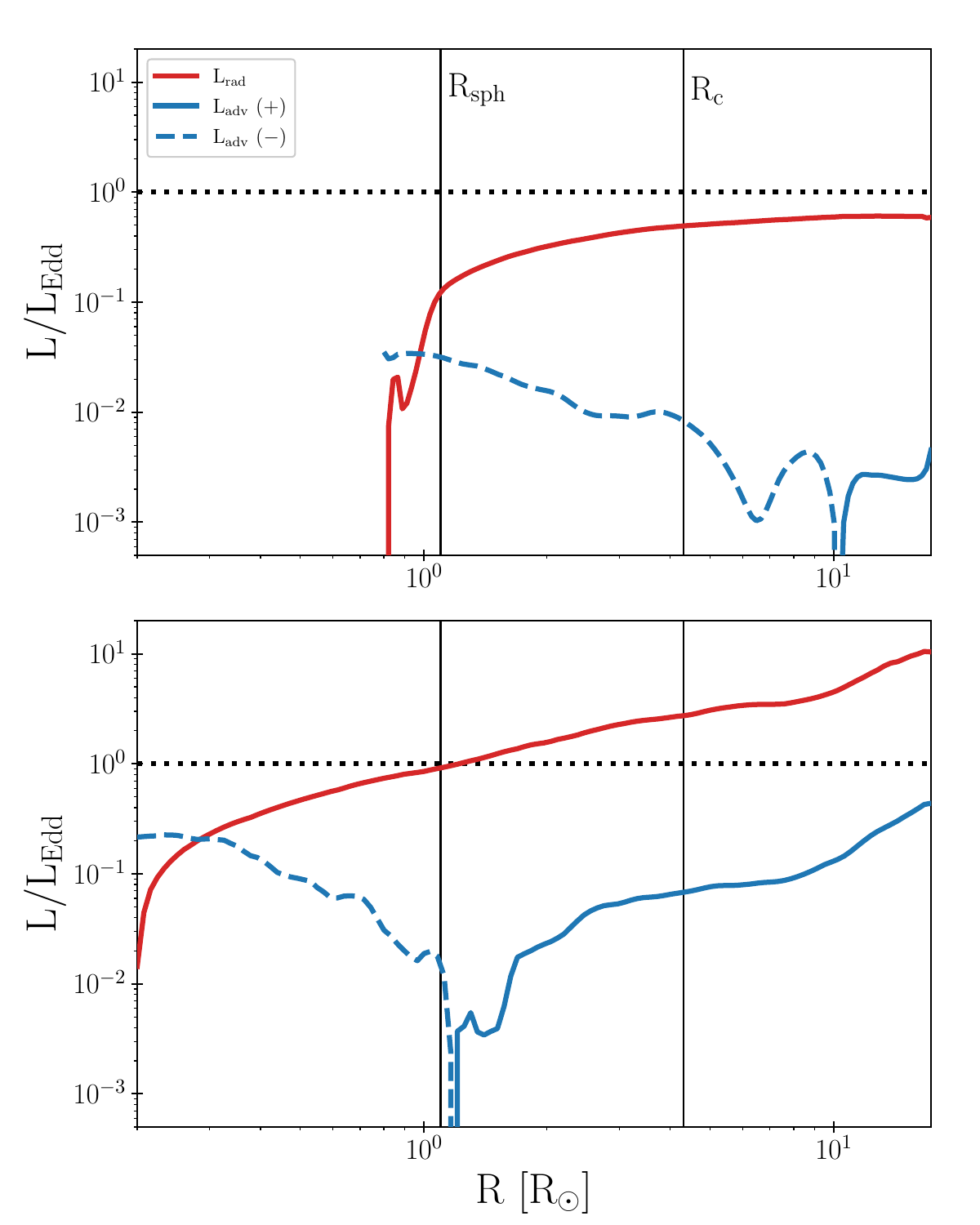}
\end{center}
\caption{
The radial profiles of two luminosities, obtained in the M3Z1\_3D (upper) and M3Z1\_L2D (lower) models.
The red and blue curves represent the radiation luminosity and the advection luminosity defined with Eq.~(\ref{eq:Lr_La}), respectively.
Here, the blue solid (dashed) curve indicates that the sign of $L_{\rm adv}$ is positive (negative), so that radiation energy advects outward (inward).
The two vertical lines denote radii corresponding to $\rsph$ and $R_{\rm c}$.
}
\label{fig:luminosity}
\end{figure}

\begin{figure}
\begin{center}
\includegraphics[width=\columnwidth]{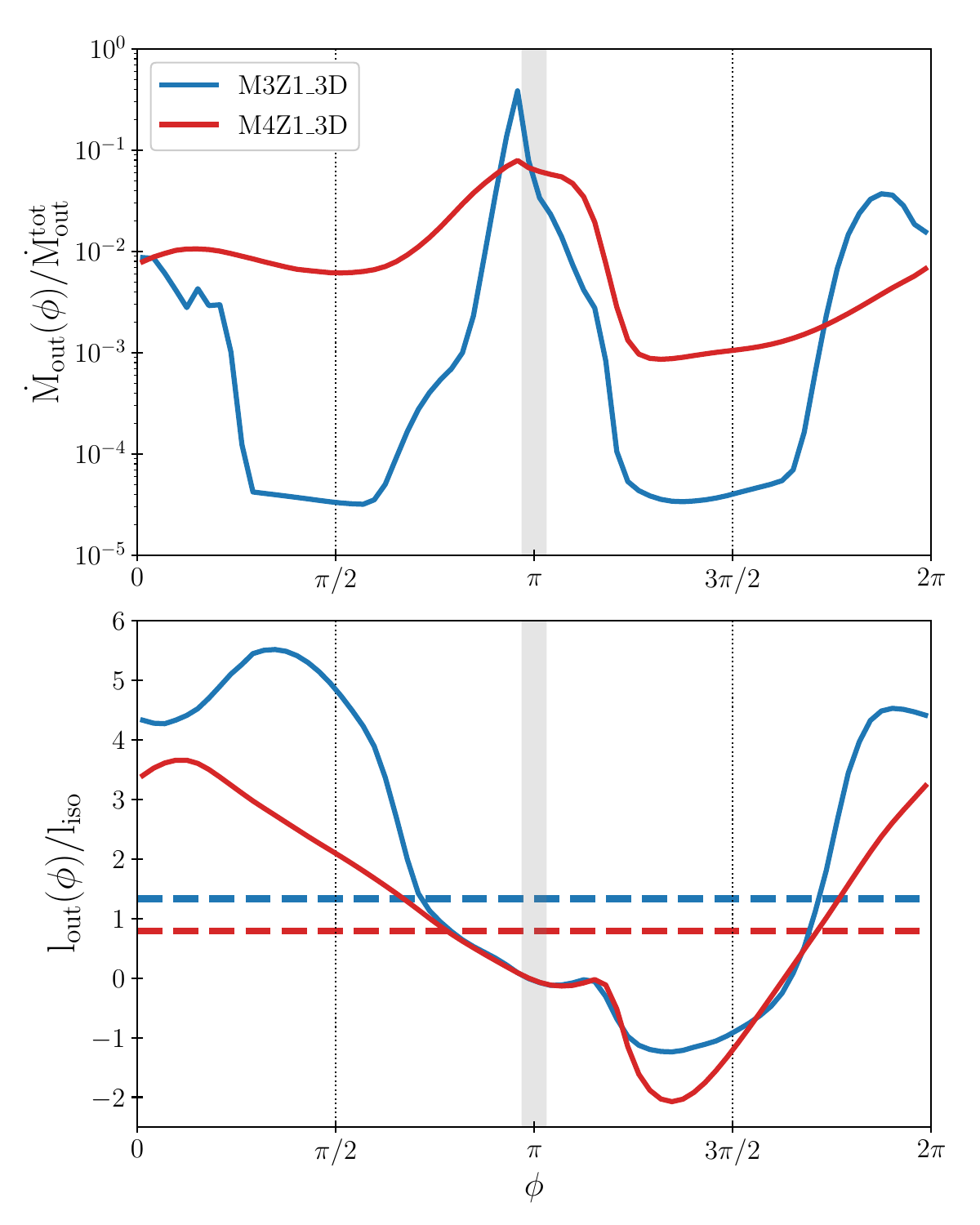}
\end{center}
\caption{
The anisotropy of outflows obtained in the M3Z1\_3D (blue) and M4Z1\_3D (red) models.
The upper panel shows the outward mass fluxes across $\rout$ for any $\phi$ directions, $\dot{M}_{\rm out}(\phi)$.
In the lower panel, the solid curves represent the specific angular momenta of outflows escaping from $\rout$ for any $\phi$ directions, and 
the dashed horizontal lines correspond to the specific angular momenta averaged over $\phi$, using Eq.~(\ref{eq:l_iso_tot}).
The grey-shaded area represents the angle where we place the L1 nozzle, $\pi-0.1 \leq \phi \leq \pi+0.1$,
and the vertical black lines denote the angles $\phi = \pi/2$ and $\pi = 3\pi/2$.
}
\label{fig:OF_phi}
\end{figure}

\section{Results}\label{sec:results}

We conduct all the simulations until the gas mass enclosed within the numerical domain converges, indicating that the accretion dynamics have reached a quasi-steady state.
We define the quasi-steady phase as $t \geq t_{\rm qs}$, as listed in Table~\ref{table:models}, and each simulation continues until reaching at least $t = 1.5~t_{\rm qs}$.
Throughout the following analysis, we focus on the quasi-steady phase, and all the physical quantities presented are averaged over this phase unless otherwise stated.
In Table~\ref{table:models}, we list, for each simulation, the inward mass flux measured at $\rin$ and the outward one at $\rout$, denoted as $\dot{M}_{\rm in}(\rin)$ and $\dot{M}_{\rm out}(\rout)$, respectively.
We find that the mass fraction removed by outflows strongly depends on both the mass transfer rate and the size of the numerical domain. 
In the case of $\mdott = 10^3~\medd$, the 2D simulations, which cover a smaller radial scale, result in the significant outflow rates of $\dot{M}_{\rm out}/\mdott \sim 1$, whereas the 3D simulation shows weaker outflows, $\dot{M}_{\rm out}/\mdott \lesssim 0.1$. 
In the case of $\mdott = 10^4~\medd$, high outflow rates of $\dot{M}_{\rm out}/\mdott \gtrsim 0.4$ are found irrespective of the size of the numerical domain. 
Additionally, we find that the difference in opacity between the solar-abundance and metal-free chemical compositions does not substantially influence the inflow and outflow rates, except for the larger values of $t_{\rm qs}$ in the metal-free cases.
In the following subsections, we further discuss these results by exploring the detailed structure of gas inflows and outflows obtained in our simulations.

\subsection{Structure of the accretion disk}\label{sec:structure}

We first discuss the structure of the accretion disk obtained from the 3D and S2D simulations, which are connected seamlessly.
The upper panels of Figure~\ref{fig:den_temp} show the distribution of gas density and gas temperature in the M3Z1\_3D and S2D models.
The gas transferred from the L1 point circularizes due to the centrifugal and Coriolis forces in the binary system, forming an accretion disk with a size of $\rc \sim 4~\rsun$.
While the transferred gas is initially as cold as $T = 10^4~\kelvin$,
the gas temperature within the disk reaches $T \sim 10^6~\kelvin$ due to viscous heating.
The accretion disk is highly convective (see the streamlines in the top-middle panel) since its optical depth is high enough to trap the diffusive photons generated near the equatorial plane.
We also note that the accretion disk becomes geometrically thicker toward the inner region,
and the scale height varies from $H/R \sim 0.2$ at $r \sim 10~\rsun$ to $H/R \sim 1$ at $r \sim 1~\rsun$.
At the inner disk region of $r < 1~\rsun$, outflows actively occur from the highly inflated disk surface.
As a result, more than 90~\% of the transferred gas is ejected from the inner disk region.
Note that the scale inside which outflows arise is generally consistent with $\rsph \sim 1~\rsun$, as shown in Eq.~(\ref{eq:R_sph}).

The lower panels of Figure~\ref{fig:den_temp} show the hydrodynamic structure obtained from the M4Z1\_3D and S2D models.
In this case, the outer disk part at $r > 1~\rsun$ significantly extends toward the equatorial direction, overfilling the Roche lobe with the transferred gas.
We note here that the density distribution along the equatorial plane is elongated toward the second quadrant $(\pi/2 \leq \phi \leq \pi)$, apparently following the ballistic trajectory shown in the left panel of Figure~\ref{fig:grid}.
This would be because the angular momentum redistribution proceeds quickly before the transferred gas sufficiently circularizes around the BH.
In fact, the M4Z1\_3D model reaches a quasi-steady state on a short timescale corresponding to the viscous time $t_{\rm v} \sim 3~\rm day$ at $r \sim R_{\rm c}$
whereas the M3Z1\_3D model takes over ten times longer to achieve it.
Interestingly, the elongated structure leads to the outflows preferentially at $\pi/2 \leq \phi \leq \pi$, as discussed in \S \ref{sec:AM_outflow}.
Moreover, the disk is highly inflated in the vertical direction, $H/R \sim 1$, even at $r \sim 10~\rsun$.
Associated with the disk inflation, dense outflows with $n_{\rm H} \gtrsim 10^5~{\rm cm^{-3}}$ arise from the outer disk region.
The presence of such large-scale outflows is a natural consequence of $\rsph > \rc$ for $\mdott = 10^4~\medd$.
On the other hand, the density and thermal structure at $r < 1~\rsun$ is generally similar to that obtained in the M3Z1 model.

The solid curves in Figure~\ref{fig:rad_profile} show the radial profiles of gas density, gas temperature, and rotational velocity obtained in the M3Z1\_3D and \_S2D models (blue) and the M4Z1\_3D and \_S2D models (red).
The values we plot here are evaluated in a density-weighted manner to trace dense accreting gas rather than rarefied outflowing gas above the disk surface.
In general, all of the three physical quantities smoothly connect between the 3D and S2D simulations.
The gas density and gas temperature in the M4Z1 model are slightly higher than the M3Z1 model while the overall profiles are common among both cases.
Interestingly, the gas density roughly follows $\rho \propto r^{-0.5}$,
which is consistent with the CDAF solution presented by \citet{Narayan2000ApJ} and \citet{Quataert2000ApJ}.
Note here that in the 3D simulations, the gas density increases outward at $r > 10~\rsun$ because the narrow and dense stream originating from the L1 nozzle dominates outside the outer edge of the disk.
Similarly, the gas temperature also increases inward as $T \propto r^{-0.5}$, in good agreement with the predictions of the slim disk model proposed by \citet{Watarai2000PASJ} and previous 2D RHD simulations \citep[e.g.,][]{Ohsuga2005ApJ, Kitaki2018PASJ}.
The rotational velocity is generally comparable to the Kepler velocity at $r \lesssim R_{\rm c}$, 
indicating that the accretion disk is primarily supported by the centrifugal force toward the radial direction.
Finally, we mention that the hydrodynamic structures obtained from the M3Z1\_L2D and M4Z1\_L2D simulations depicted by the dashed curves in Figure~\ref{fig:rad_profile} are generally consistent with those from the 3D and S2D simulations in spite of the difference in the grid configuration and boundary conditions.

\subsection{Generation and propagation of outflows}\label{sec:outflow}

\subsubsection{3D and S2D simulations}

As in \S \ref{sec:structure}, we first focus on the results of the 3D and S2D simulations.
The solid curves in Figure~\ref{fig:mass_flux} show the inward and outward mass fluxes at any radii obtained from the M3Z1\_3D and \_S2D models (upper) and the M4Z1\_3D and \_S2D models (lower).
We find that both models have similar profiles in mass fluxes.
The inward and outward mass fluxes have a peak at $r \gtrsim R_{\rm c}$,
indicating the accumulation of transferred gas at the outer disk edge. 
Inside the disk, the inward mass flux decreases for smaller radii 
as a result of the gradual mass loss associated with outflows.
The radial profile approximately follows $\dot{M}_{\rm in} \propto r^{0.6}$, in good agreement with the recent 2D RHD simulations by \citet{Hu2022ApJ}.
Note here that this slope of the inward mass fluxes is shallower than that predicted by the windy disk model of \citet{Shakura1973A&A}. 
The windy disk model requires $\dot{M}_{\rm in} \propto r$, such that the local energy flux generated by the disk's viscosity equals the local Eddington flux at all radii.
We consider that the shallower mass flux profile in our simulations is attributed to effective cooling by advection of photons, 
which is included in our simulations but not considered in the windy model \citep[e.g.,][]{Abramowicz1988ApJ, Ohsuga2005ApJ, Ohsuga2011ApJ, Sadowski2009ApJS, Sadowski2011A&A}.

To understand the nature of outflows, we explore the energetics of the outflowing gas, using the Bernoulli number defined in the co-rotating frame as
\begin{eqnarray}
Be = \frac{1}{2} v^2 + \frac{5k_{\rm B} T}{2 \mu_{\rm g} m_{\rm p}} + \frac{4 \erad}{3\rho} + \Phi(\rvec) \ ,
\label{eq:Be}
\end{eqnarray}
where the Roche potential is given by
\begin{eqnarray}
\frac{-a\Phi(\rvec)}{G M_{\rm tot}} = \frac{a\mu}{|\rvec|} + \frac{a(1-\mu)}{|\rvec - \rvec_\ast|} + \frac{(x-x_{\rm b})^2 + y^2}{2a^2} \ 
\label{eq:Phi_be}
\end{eqnarray}
\citep[e.g.,][]{Shu1979ApJ, Lu2023MNRAS}{}{}.
The Bernoulli number indicates how far outflows potentially propagate from the central BH.
In general, outflows with $Be > \Phi_{\rm L2}$, where $\Phi_{\rm L2}$ is the Roche potential at the L2 point, flow away from the Roche lobe and consequently extract masses and angular momenta from the binary. 

Here, we evaluate the Bernoulli number of outflowing gas as a function of radius.
The left column of Figure~\ref{fig:mf_Be} shows $\dot{M}_{\rm out}(r)$ evaluated for the five bins of $Be$, obtained in the M3Z1\_3D and \_S2D models (top) and the M4Z1\_3D and \_S2D models (bottom).
In both cases, we find a gradual increase in the typical values of $Be$ for outflowing gas with increasing radii.
Outflows occupying small radii $r < \rc$ are typically characterized by low Bernoulli numbers $Be < \Phi(\rc)$,
indicating that they remain strongly bound within the binary system.
Conversely, the outskirts of the accretion disk are occupied by outflows with higher Bernoulli numbers.
Specifically, in the M4Z1\_3D model, marginally unbound outflows characterized by $\Phi_{\rm L2} < Be \leq 0$ account for over 90~\% of the outflow rates across $\rout$,
whereas the M3Z1\_3D model hardly produces them.
We also find that while unbound outflows with $Be > 0$ can arise from various radii owing to intense acceleration caused by Thomson scattering with X-ray irradiation, their outflow rates are usually small.

It is worth noting here that the outflow rates shown in Figure~\ref{fig:mf_Be} contain contributions from both outward convective motions within the accretion disk and disk winds arising from the disk surface. 
Primarily, the weak outflows observed in the inner disk regions are associated with convective motions,
while the marginally unbound outflows emerging from the outer disk edge are typically regarded as disk winds.
Based on these findings, we claim that gas lifted by convective motions at an inner radius often falls back on a slightly outer radius, where it undergoes further acceleration due to diffusive radiation pressure, subsequently propelling it outward with higher $Be$.
As a result, gas transported outward due to the weak outflow accumulates at the outskirts of the disk until it overflows from the binary potential.

However, we are concerned that the 3D simulations likely underestimate the outflow rates across $\rout$ because the radius of the inner boundary, $\rin$, is too large to track the outward transport of gas from $r < R_{\rm sph}$.
In the following, we address this concern using the L2D simulations, where $\rin$ is 4 times smaller than in the 3D simulations.

\subsubsection{L2D simulations}

The dashed curves in Figure~\ref{fig:mass_flux} show the inflow and outflow profiles in the M3Z1\_L2D and M4Z1\_L2D models.
The overall profiles are similar to those in the 3D and S2D simulations:
both inflow and outflow rates exhibit peaks at $r \gtrsim R_{\rm c}$ and 
gradually decline inward following $\dot{M} \propto r^{0.6}$ at $r \lesssim 1~\rsun$ for the M3Z1\_L2D model and at $r \lesssim 3~\rsun$ for the M4Z1\_L2D model.
A remarkable difference from the 3D simulations is that the M3Z1\_L2D and M4Z1\_L2D models have considerably high outflow rates near the outermost radius, $\dot{M}_{\rm out}/\mdott \sim 0.5$ and $0.9$, respectively.
This implies that the smaller innermost radius in the L2D simulations leads to more efficient acceleration by radiation pressure in the disk outskirts.
A similar trend is found in the $Be$ distribution of outflows, which is shown in the right column of Figure~\ref{fig:mf_Be}.
$\dot{M}_{\rm out}(r,~Be)$ in the L2D simulations are similar to those in the 3D and S2D simulations, except for the outermost regions, where the outflow rates are clearly dominated by high-energy outflows with $Be > \Phi_{\rm L2}$.
Thus, the L2D simulations provide substantial outflows that are energetically strong enough to leave the binary.

We here explore how those powerful outflows occur in the L2D simulations.
Figure~\ref{fig:den_mf_25D} shows the time evolution of the radiation temperature distribution, denoted as $T_{\rm r} \equiv (E_{\rm r}/a_{\rm r})^{1/4}$, and the mass flux distribution, obtained in the M3Z1\_L2D model. 
This figure indicates that a hot convective region with $T_{\rm r} \sim 10^6~\kelvin$ is developed at the outskirt of the disk and gradually extends outward by accumulating inflowing gas from the outer boundary and outflowing gas expelled from the inner disk region.
This extension of the convective regime finally produces large-scale outflows across the outer boundary.
Note that while such a convective region also appears in the M3Z1\_3D model, 
its growth ceases without driving large-scale outflows.
This difference between the L2D and 3D simulations arises from the different inner boundaries, which change the whole energy released by gas accretion and the nature of its propagation.
Figure~\ref{fig:luminosity} shows the radial profiles of the radiation luminosity and the advection luminosity obtained in the M3Z1\_3D and M3Z1\_L2D models, which are evaluated as follows:
\begin{eqnarray}
L_{\rm rad} = \int \Fvec \cdot {\rm d} \mbox{\boldmath $S$},~{\rm and}~
L_{\rm adv} = \int \erad \vvec \cdot {\rm d} \mbox{\boldmath $S$},
\label{eq:Lr_La}
\end{eqnarray}
where ${\rm d} \mbox{\boldmath $S$} = r^2{\rm sin}\theta {\rm d}\theta{\rm d}\phi \evec_{r}$.
This figure shows that the M3Z1\_3D model provides the sub-Eddington radiation luminosity,
and the advection luminosity is always negative inside $\rc$, corresponding to net inward energy fluxes in the accretion disk.
On the other hand, the M3Z1\_L2D model provides the super-Eddington radiation luminosity at $r > \rc$,
which is several times higher than that obtained in the M3Z1\_3D model.
This is naturally explained because the radiation luminosity generated by viscous heating within the numerical domain is related to the innermost radius as $L_{\rm rad} \sim G \mbh \mdott / \rin$.
In addition, the advection luminosity becomes positive outside $\rsph$,
which indicates that the radiation energy is transported by outflowing gas.
The outward advection luminosity gradually increases outside $\rc$, reaching about 50~\% of the Eddington luminosity around $\rout$.
Thus, the M3Z1\_L2D model produces more energy released by gas accretion than the M3Z1\_3D model and efficiently transfers it outward by radiation and advection.
Based on these facts, we claim that the outward propagation of energy and momentum feeds the convective region in the disk outskirt, finally driving substantial mass loss from the binary.

Here, it is worth mentioning that the discrepancy in the outflow rates between the L2D and 3D models is not likely due to the difference in dimensions.
In the 3D simulations, angular momentum transport and viscous heating are also driven by the Reynolds stress associated with dynamical turbulence within the accretion disk as well as the Maxwell stress included as the explicit viscosity in this study.
To check this effect, we evaluated, for the M3Z1\_3D model, the viscous parameter originating from the Reynolds stress as
$\alpha_{\rm R}(r) = (2/3) \left < \rho \delta v_{r} \delta v_\phi \right > /\left < p \right >$, where angle brackets denote taking the average for any quantity over the $\theta$ and $\phi$ directions at any radius $r$, and $\delta v_{i} \equiv v_{i}(r,\theta,\phi) - \left < v_{i}(r) \right >$.
Consequently, we found that $\alpha_{\rm R}(r) \sim 10^{-3}$ at $r < \rc$, about one hundred times smaller than 
$\alpha = 0.1$, which is assumed to mimic the Maxwell stress in our simulations.
Therefore, we consider that the three-dimensional turbulent motions within the disk provide a minor effect on the significant difference in the observed outflow structure between the 3D and L2D simulations.

\subsection{Angular momentum of outflows}\label{sec:AM_outflow}

Although we mentioned that our 3D simulations underestimate outflow rates,
they are still helpful for learning the anisotropy in outflows and the angular momenta removed from the binary.
The top panel of Figure~\ref{fig:OF_phi} shows $\dot{M}_{\rm out}(\rout)$ at any azimuthal direction $\phi$ obtained from the M3Z1\_3D model (blue) and the M4Z1\_3D model (red).
We find that both models have two peaks in the mass loss rates.
The first peak emerges around $\phi = 0$ and $2\pi$, the furthest point from the donor, where the strong centrifugal force in the binary accelerates gas outward.
The second peak appears around the $\phi = \pi$, where we place the L1 nozzle (grey-shaded). 
This indicates that the dense stream coming through the L1 nozzle loses some fraction of mass before reaching the accretion disk,
and we consider that the mass loss is caused by the high gas pressure within the stream and the local minimum in the Roche potential at the L1 point.
We also note that the outflow rate in the M4Z1 model is dominated by outflows from the second quadrant.
This would reflect that the accretion disk is rather elongated toward $\phi \sim 3\pi/4$, following the streaming line of transferred material after the first pericentric passage, as found in Figure~\ref{fig:den_temp}.
Nevertheless, the anisotropy of the outflow rates in the M4Z1 model is more modest than in the M3Z1 model.
This is because, under a rapid mass transfer with $\mdott = 10^4~\medd$,
the radiation pressure in the accretion disk becomes strong enough to drive outflows toward any $\phi$ direction.


Next, we discuss the angular momentum that outflows potentially carry away from the binary.
Based on our 3D simulations, we evaluate the specific angular momentum of outflows $l_{\rm out}$ relative to the binary's barycenter.
The bottom panel of Figure~\ref{fig:OF_phi} shows $l_{\rm out}$ for each $\phi$ direction.
Here, the plotted value of $l_{\rm out}$ is normalized by the specific angular momentum expected in the isotropic emission case, where outflows blow isotropically in the BH's rest flame, which is described as follows:
\begin{eqnarray}
l_{\rm iso} = \frac{\ms^2}{M_{\rm tot}^2} \sqrt{ G M_{\rm tot} a} \ .
\label{eq:l_iso}
\end{eqnarray}
This figure indicates that $l_{\rm out}$ reaches the peak around $\phi = 0$ and $2\pi$, where the rotational velocity relative to the binary's barycenter becomes maximum.
Apart from these angles, $l_{\rm out}$ decreases and reaches the bottom in the third quadrant, where the Coriolis force accelerates gas in the opposite direction relative to the binary's rotation.
Then, we evaluate the specific angular momenta averaged over the entire outflows as follows:
\begin{eqnarray}
l^{\rm avg}_{\rm out} = 
\frac{\int l_{\rm out} \dot{M}_{\rm out} {\rm d}\phi}
{\int \dot{M}_{\rm out} {\rm d}\phi} \ .
\label{eq:l_iso_tot}
\end{eqnarray}
The thick dashed lines in Figure~\ref{fig:OF_phi} represent the values of $l^{\rm avg}_{\rm out}/l_{\rm iso}$, which are 1.33 and 0.78 for the M3Z1\_3D and M4Z1\_3D models, respectively (see also Table~\ref{table:models}),
suggesting that the specific angular momenta obtained in both models are roughly identical to that in the isotropic emission case, in spite of the anisotropy in the outflow structure.

Finally, we note that the M3Z0\_3D and M4Z0\_3D models, which assume the metal-free opacity, yield $l^{\rm avg}_{\rm out}/l_{\rm iso} = 3.21$ and $0.30$, respectively, which are apart from unity, compared to the solar abundance models.
This suggests that the absence of heavy elements promotes the anisotropy of outflows.
Specifically, the M3Z0\_3D and M4Z0\_3D models produce more outflows biased to $\phi \sim 0$ and $\phi \sim 3\pi/4$, respectively, compared to their solar-abundance counterparts.
This discrepancy can be attributed to the reduction of radiative pressure in the accretion disk due to the absence of heavy elements, which aids in maintaining anisotropic structures.

\subsection{Metallicity-dependence of the disk evolution}\label{sec:metal}

\begin{figure}
\begin{center}
\includegraphics[width=0.95\columnwidth]{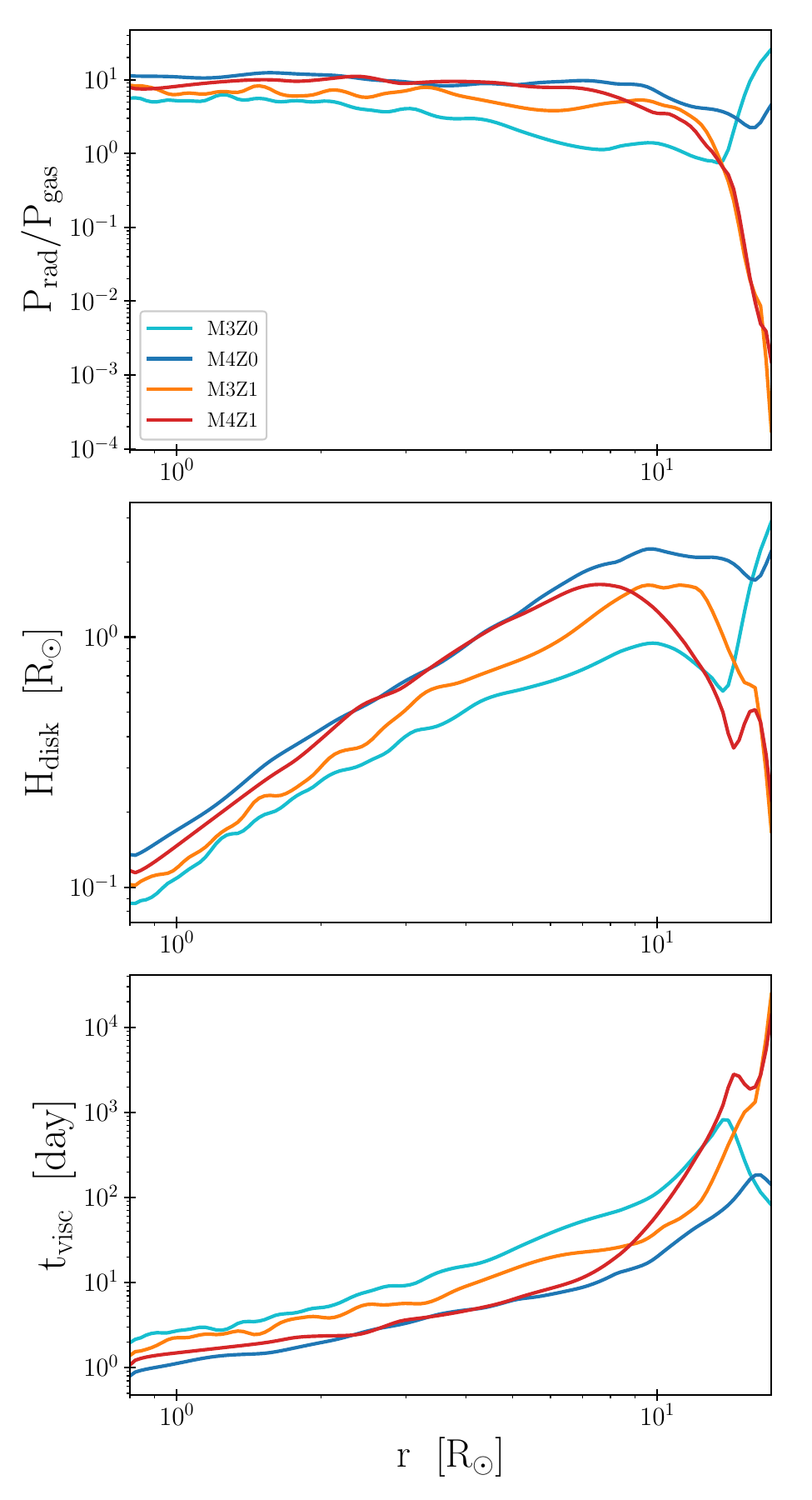}
\end{center}
\vspace{-5mm}
\caption{
The influence of heavy elements on the accretion dynamics.
From top to bottom, each panel displays the radial profiles of the radiation-to-gas pressure ratio, the disk thickness, and the viscous time, obtained in the 3D simulations.
The cyan and blue curves represent the M3Z0 and M4Z0 models, respectively, using the metal-free opacity.
On the other hand, the orange and red curves represent the M3Z1 and M4Z1 models, respectively, using the solar-abundance opacity.
}
\label{fig:P_tadv}
\end{figure}

Table~\ref{table:models} shows that 
the models with the metal-free opacity do not provide a significant difference in $\IFR$ and $\OFR$ from their companion models assuming the solar-abundance opacity.
However, as found from $t_{\rm qs}$ listed in Table~\ref{table:models},
the metal-free models require a longer time to reach a quasi-steady state compared to the solar-abundance model,
as especially pronounced in the comparison between the M3Z0\_3D and M3Z1\_3D models.
This difference in $t_{\rm qs}$ implies that the presence of heavy elements affects the dynamical evolution of the accretion disk.
In Figure~\ref{fig:P_tadv}, we compare the four 3D simulations in terms of the radial profiles of the radiation-to-gas pressure ratio, the thickness of the accretion disk, and the viscous timescale.
Here, the disk thickness and the viscous timescale are evaluated as $H_{\rm d} = c_{\rm s}/\Omega$ and $t_{\rm v} = (r/H_{\rm d})^2/(\alpha \Omega)$, respectively.
The comparison between the M3Z0 and the M3Z1 models demonstrates that 
the solar-abundance opacity enhances the radiation-to-gas pressure ratio by a factor of a few compared to the metal-free opacity.
This enhanced radiation pressure makes the accretion disk geometrically thicker and consequently leads to a shorter viscous time:
the M3Z1 (M3Z0) model provides $t_{\rm v} \sim 8~(15)~{\rm days}$ around $r \sim R_{\rm c}$, implying $t_{\rm qs}~\sim~5 t_{\rm v}(R_{\rm c})$.
Note here that the evaluated viscous timescales in the M4Z0 and M4Z1 models are roughly comparable to each other, consistent with the small difference in $t_{\rm qs}$.
Thus, while radiation pressure enhancement by the presence of heavy elements provides no significant impact on the generation of outflows,
it effectively enhances viscosity in the disk and thus influences gas accretion onto the central BH.

\section{Discussion}\label{sec:discussion}

\renewcommand{\arraystretch}{1.5}
\begin{table*}
\begin{center}
\begin{tabular}[c]{ccccccccc} \hline \hline
Reference & Method & $\mbh$ & $\rin$ & $\rout$ & $R_{\rm c}$ & $\dot{M}_{\rm in}(\rin)$ & $\dot{M}_{\rm t}$ & $R_{\rm sph}/\rout$ \\
 & & $[\msun]$ & $[r_{\rm S}]$ & $[r_{\rm S}]$ & $[r_{\rm S}]$ & $[\medd]$ & $[\medd]$ & \\ \hline
M3Z1\_S2D+3D & 2D+3D-RHD & 34 & $\sim 140$  & $\sim 10^{5}$ & $\sim 30000$ & 120  & 1000  & $\sim 0.06$   \\
M3Z1\_L2D    & 2D-RHD    & 34 & $\sim 1400$ & $\sim 10^{5}$ & $\sim 30000$ & 720  & 1000  & $\sim 0.06$   \\
M4Z1\_S2D+3D & 2D+3D-RHD & 34 & $\sim 140$  & $\sim 10^{5}$ & $\sim 30000$ & 590  & 10000 & $\sim 0.6$  \\ 
M4Z1\_L2D    & 2D-RHD    & 34 & $\sim 1400$ & $\sim 10^{5}$ & $\sim 30000$ & 1500 & 10000 & $\sim 0.6$  \\ 
\hline
Jiang+14 & 3D-RMHD & 6.62 & 2 & 50 & 25 & $\sim 22$ & $\sim 62$  & $\sim 9$ \\
Hashizume+15 & 2D-RHD & 10 & 3 & 5000 & 100 & $\sim 15$ & $100$  & 0.15   \\
             &        & 10 & 3 & 5000 & 500 & $\sim 10$ & $100$  & 0.15   \\
             &        & 10 & 3 & 5000 & 100 & $\sim 70$ & $1000$ & 1.5    \\
S\c{a}dowski+15 & 2D-GR-RMHD & 10 & 0.925 & 2500 & 21 & $\sim 42$ & $\sim 740$ & $\sim 2$ \\
S\c{a}dowski+16 & 3D-GR-RMHD & 10 & 0.925 & 500  & 20 & $\sim 18$ & $\sim 70$  & $\sim 1$  \\
Kitaki+18   & 2D-RHD & 10 & 2 & 3000 & 100 & $\sim 14$  & $30$    & 0.075 \\
            &        & 10 & 2 & 3000 & 300 & $\sim 28$  & $100$   & 0.25  \\
            &        & 10 & 2 & 3000 & 300 & $\sim 79$  & $500$   & 1.25  \\
            &        & 10 & 2 & 3000 & 300 & $\sim 83$  & $1000$  & 2.5   \\
            &        & 10 & 2 & 3000 & 300 & $\sim 550$ & $10000$ & 25    \\
Kitaki+22   & 2D-RHD & 10 & 2 & 3000 & 2430 & $\sim 18$ & $70$    & 0.175 \\
Yoshioka+22 & 2D-RHD & 10 & 2 & 6000 & 2430 & $\sim 11$ & $35$    & 0.04  \\
            &        & 10 & 2 & 6000 & 2430 & $\sim 13$ & $50$    & 0.06  \\
            &        & 10 & 2 & 6000 & 2430 & $\sim 14$ & $70$    & 0.09  \\
            &        & 10 & 2 & 6000 & 2430 & $\sim 18$ & $70$    & 0.09  \\
            &        & 10 & 2 & 6000 & 2430 & $\sim 38$ & $200$   & 0.25  \\
Hu+22       & 2D-RHD & 1000 & 3 & 1500 & 1250 & $\sim 88$ & $1408$ & $\sim 7$ \\
\hline \hline \\
\end{tabular}
\end{center}
\caption{
Comparison of our study with previous multi-dimensional RHD simulations \citep[][]{Jiang2014ApJ, Hashizume2015PASJ, Sadowski2015MNRAS, Sadowski2016MNRAS, Kitaki2018PASJ, Kitaki2021PASJ, Yoshioka2022PASJ, Hu2022ApJ}. 
The reference papers are taken from Table~1 of \citet{Kitaki2021PASJ}, except for the addition of \citet{Yoshioka2022PASJ} and \citet{Hu2022ApJ}.
The first four lines present our models while the subsequent ones summarize the models provided by the reference papers. 
The models labeled "S2D+3D" represent numerical setups and results combined from our S2D and 3D simulations that consistently connect the inward mass fluxes at the boundary.
Specifically, the listed values of $\rin$ and $\dot{M}_{\rm in}(\rin)$ are based on the S2D simulations, whereas other quantities are based on the 3D simulations.
The second column indicates the simulation method used for each reference.
"RMHD" denotes radiation magnetohydrodynamical simulations, while "GR" corresponds to general relativistic simulations.
From the third to sixth columns, we present, for each reference, BH mass, inner and outermost radii, and centrifugal radius of accreting gas.
The seventh column indicates the inward mass flux at the innermost radius, which is identical to the mass accretion rate onto the BH, except for our simulations, where $\rin$ is much larger than the ISCO radius.
The eighth column denotes the mass injection rate from the outer boundary, which corresponds to the mass transfer rate in our simulation.
For \citet{Jiang2014ApJ}, \citet{Sadowski2015MNRAS}, and \citet{Sadowski2016MNRAS}, which suppose a gas torus as the initial condition rather than continuous gas supply from the outer boundary, we evaluate $\mdott \equiv \dot{M}_{\rm in}(\rin)+\dot{M}_{\rm out}(\rout)$.
The ninth column presents the expected spherization radius normalized by the outermost radius,
which is calculated by substituting $\mdott$ into Eq.~(\ref{eq:R_sph}).
}
\label{table:compare}
\end{table*}
\renewcommand{\arraystretch}{1}

\begin{figure}
\begin{center}
\includegraphics[width=\columnwidth]{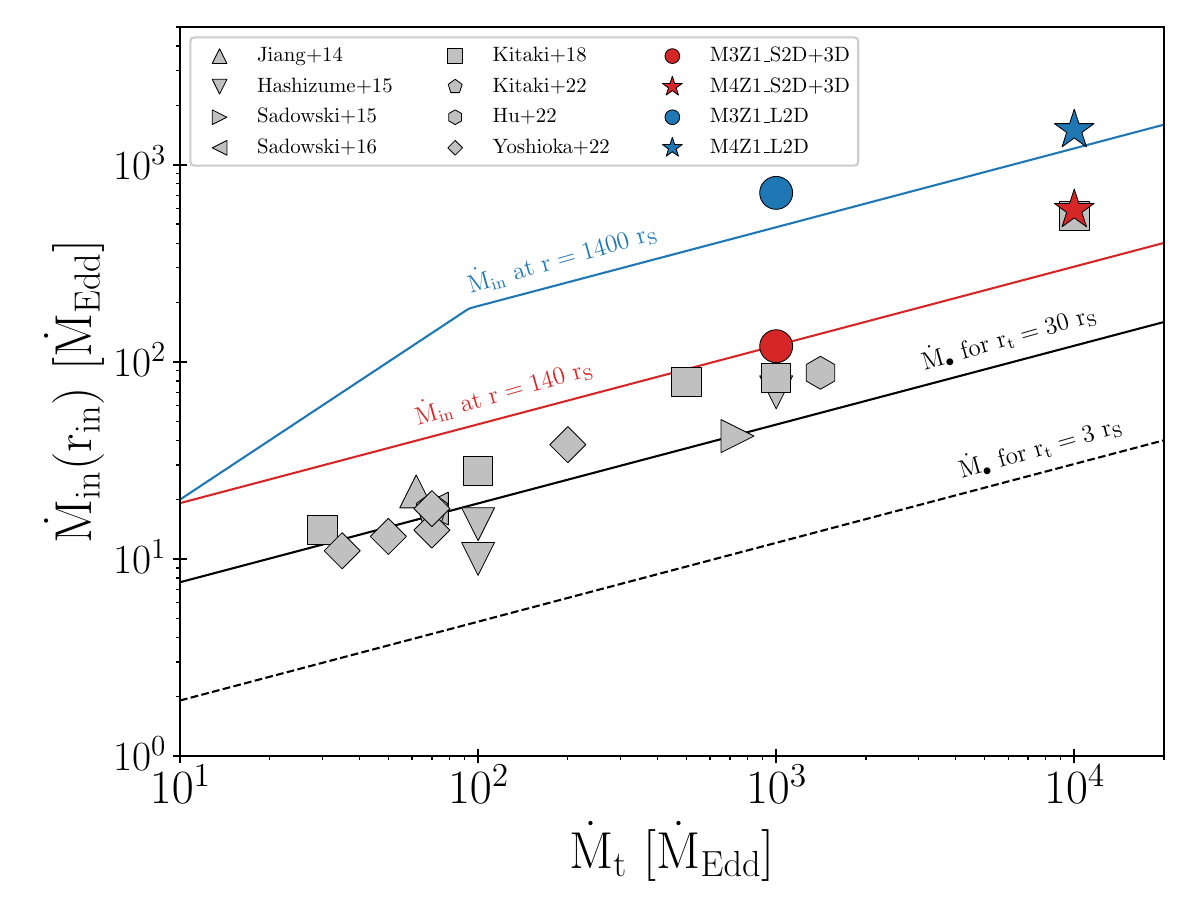}
\end{center}
\vspace{-5mm}
\caption{
The comparison of our simulations with previous multi-dimensional RHD simulations (the reference models are listed in Table~\ref{table:compare}).
The values in the x-axis and the y-axis represent $\mdott$ and $\dot{M}_{\rm in}(\rin)$, respectively.
The blue circle and star denote the results of the M3Z1\_L2D and  M4Z1\_L2D models with $\rin \sim 1400~r_{\rm S}$,
whereas the red circle and star denote the results of the M3Z1\_S2D+3D and  M4Z1\_S2D+3D models with $\rin \sim 140~r_{\rm S}$.
For comparison, the blue and red solid lines indicate $\dot{M}_{\rm in}$ evaluated with Eq.~(\ref{eq:mdotin}) for $r = 1400~r_{\rm S}$ and $140~r_{\rm S}$, respectively.
The grey symbols represent the results obtained in the previous simulations, in which the values of $\dot{M}_{\rm in}(\rin)$ generally correspond to the mass accretion rates on the central BH.
For comparison with those plots, the dotted and solid black lines show $\dot{M}_\bullet,3$ and $\dot{M}_\bullet,30$, respectively (see \S~\ref{sec:acc_rate} for the definitions).
This figure indicates that the formula given by Eq.~(\ref{eq:mdotin}) self-consistently explains $\dot{M}_{\rm in}(\rin)$ in our simulations and the reference papers, particularly when the reduction of $\dot{M}_{\rm in}$ terminates around a few tens times $r_{\rm S}$.
}
\label{fig:comp}
\end{figure}

\subsection{Mass accretion rates onto the central BH}\label{sec:acc_rate}

Our 2D and 3D RHD simulations unveiled the quasi-steady hydrodynamic structure established during super-Eddington mass transfer to a stellar-mass BH in a binary system, particularly focusing on the outer radii $r \sim 100$--$10^5~r_{\rm S}$.
We revealed that at the outskirts of the accretion disk, mass fluxes are enhanced by a factor of a few to the mass transfer rate due to the accumulation of inward and outward flows,
whereas, inside the spherization radii, the inward mass fluxes gradually decline toward smaller radii, following $\dot{M}_{\rm in} \propto r^{0.6}$.
Based on these findings, we propose an approximate formula for the inward mass flux as a function of any $\mdott$ and $r$:
\begin{eqnarray}
\dot{M}_{\rm in} \approx 
\begin{cases}
~\dot{M}_{\rm t}^{\prime} & (r > R_{\rm sph}^{\prime}) \\
~\dot{M}_{\rm t}^{\prime} ( r / R_{\rm sph}^{\prime} )^{0.6} & (r_{\rm t} \leq r < R_{\rm sph}^{\prime}) \\
~\dot{M}_{\rm in}(r_{\rm t}) & (r < r_{\rm t}) \ \ ,
\end{cases}
\label{eq:mdotin}
\end{eqnarray}
where we adopt $\dot{M}_{\rm t}^{\prime} = 2~\mdott$ and $R_{\rm sph}^{\prime} = 7.5 (\dot{M}_{\rm t}^{\prime}/\medd) r_{\rm S}$ to account for gas accumulation at the outskirts of the disk.
In Figure~\ref{fig:comp}, we compare $\dot{M}_{\rm in}(\rin)$ obtained in our simulations with the corresponding values given by Eq.~(\ref{eq:mdotin}) (see the blue and red symbols and lines).
This comparison demonstrates that the approximate formula generally reproduces our simulation results within at most $\sim 50~\%$ difference.

In Eq.~(\ref{eq:mdotin}), we suppose that the power-law decrease of $\dot{M}_{\rm in}$ terminates at $r = r_{\rm t}$, 
thereby the mass accretion rate on the central BH is written as $\dot{M}_\bullet = \dot{M}_{\rm in}(r_{\rm t})$.
A conservative limit in the accretion rate is set with $r_{\rm t} = 3~r_{\rm S}$, corresponding to the innermost stable circular orbit (ISCO) radius for a non-spinning BH. 
In this case, Eq.~(\ref{eq:mdotin}) yields $\dot{M}_{\rm \bullet} \sim 12~(30)~\medd$ for $\mdott \sim 10^3~(10^4)~\medd$ (hereafter denoted as $\dot{M}_{\rm \bullet,3}$).
On the other hand, several RHD simulations that followed super-Eddington accretion flows down to the ISCO radius suggest that $r_{\rm t}$ is several times larger than the ISCO radius,
inside which the net inward energy flux is realized due to photon trapping wihtin the accretion disk \citep[e.g.,][]{Hashizume2015PASJ, Sadowski2015MNRAS, Sadowski2016MNRAS, Kitaki2018PASJ, Kitaki2021PASJ}.
By applying $r_{\rm t} = 30~r_{\rm S}$ for Eq.~(\ref{eq:mdotin}), one obtaines $\dot{M}_{\rm \bullet} \sim 48~(120)~\medd$ for $\mdott \sim 10^3~(10^4)~\medd$ (hereafter denoted as $\dot{M}_{\rm \bullet,30}$).

Figure~\ref{fig:comp} presents $\dot{M}_\bullet$ evaluated with Eq.~(\ref{eq:mdotin}) as well as those directly measured in multi-dimensional RHD simulations in the literature (see Table~\ref{table:compare} for the references).
These simulations show a positive correlation between $\dot{M}_{\rm \bullet}$ and $\mdott$, irrespective of variations in BH mass, centrifugal radius of accreting gas, and numerical methods adopted in each simulation.
The slope and normalization in this correlation are generally consistent with those exhibited by $\dot{M}_{\rm \bullet,30}$,
while $\dot{M}_{\rm \bullet,3}$ is clearly lower than the simulation data.
Thus, the power-law dependence of $\dot{M}_{\rm in} \propto r^{0.6}$, starting nearly from $R_{\rm sph}$ and ceasing around a few tens times $r_{\rm S}$, nicely explains an empirical characteristic of super-Eddington accretion disks undergoing radiation-driven mass loss.

However, caution is warranted when comparing Eq.~(\ref{eq:mdotin}) to simulations with a numerical domain that is smaller than the spherization radii.
For instance, \cite{Kitaki2018PASJ} find $\dot{M}_\bullet \sim 550~\medd$ in the case of $\mdott = 10^4~\medd$, which is approximately five times higher than the expectation of $\dot{M}_{\rm \bullet,30}$ (indicated by the rightmost square behind the red star in Figure~\ref{fig:comp}).  
However, they adopt the outermost radius of the numerical domain as $R_{\rm sph}/\rout = 25$. 
This suggests that the inward mass flux reaching the central BH is overestimated by a factor of $(R_{\rm sph}/\rout)^{0.6} \sim 7$ because the simulation effectively limits the radii generating outflows.
On the other hand, RHD simulations by \citet{Jiang2014ApJ} and \citet{Hu2022ApJ}, which also adopted small numerical domains with $R_{\rm sph}/\rout \sim 9$ and $7$, respectively, yielded the mass accretion rates roughly comparable to $\dot{M}_{\rm \bullet,30}$.
These simulations showed that the inward mass fluxes remain decreasing down to the vicinity of the central BH by outflows, so that $r_{\rm t} \sim 3~r_{\rm S}$ rather than $30~r_{\rm S}$ can compensate the overestimation in $\dot{M}_{\rm in}$ by a factor of $(R_{\rm sph}/\rout)^{0.6} \sim 4$.
Thus, the critical radius to terminate outflows may vary depending on gas inflow rates at larger radii, influenced by the location of the spherization radius relative to the outer boundary in the numerical domain.
It is worth emphasizing here that our 2D and 3D RHD simulations, which considered the outer boundary physically motivated in a mass-transferring binary, highlight those outer boundary effects that previous smaller-scale simulations would have suffered.

Finally, we discuss mass loss rates via radiation-driven outflows from binaries, based on the above argument.
When the entire hydrodynamic structure is in a quasi-steady state, the outflow rates counted with gas reaching infinity must be $\dot{M}_{\rm out} = \mdott - \dot{M}_{\rm \bullet}$.
Then, if we describe the mass accretion rate on the central BH as $\dot{M}_{\rm \bullet,30}$, 
the fraction of the outflow to mass transfer rates becomes $\dot{M}_{\rm out}/\mdott \sim 0.8$ and $0.99$ for $\mdott = 100~\medd$ and $10^4~\medd$, respectively.
Thus, during super-Eddington mass transfer to a stellar-mass BH, a huge fraction of the transferred material is expelled from the binary system.
Such a substantial mass loss can lead to efficient extraction of the binary's angular momentum, as discussed in more detail in the next subsection.

\subsection{Destination of outflows}\label{sec:fate_outflows}


Here, we discuss the destination of outflows found in our RHD simulations.
As shown in \S \ref{sec:outflow}, the major fraction of outflows has Bernoulli numbers $\Phi_{L2} \leq Be < 0$, implying that they are still bound in the binary.
Such marginally bound outflows are expected to leak out from the L2 point and form a toroidal structure rotating around the binary \citep[e.g.,][]{Lu2023MNRAS}.
The circum-binary torus can gradually extend and lose its mass by acquiring energy from the binary's tidal torque.
\citet{Shu1979ApJ} has calculated the evolution of motion of a pressure-less particle, which initially corotates with the L2 point, 
and showed that it finally becomes unbound in the case of $0.06 \lesssim q~({\rm or}~q^{-1}) \lesssim 0.8$ (here $q \sim 0.83$ in our simulations).
Additionally, \citet{Pejcha2016MNRAS} has demonstrated that thermal and radiation pressure aid in driving the circum-binary material unbound.
Based on these facts, we expect that outflows with $\Phi_{\rm L2} \leq Be < 0$ found in our simulations finally become unbound and extract mass and angular momenta from the binary.

In this context, our simulations that do not follow the further acceleration outside the Roche lobe can underestimate the angular momentum carried away by outflows.
According to \citet{Shu1979ApJ}, gas reaching infinity from a binary possess $l_{\rm out} \sim 1.1$-$1.2~l_{\rm L2}$,
where $l_{\rm L2} \approx 1.44~\sqrt{G M_{\rm tot} a}$ is the specific angular momentum of a particle corotating with the L2 point \citep{Pribulla1998CoSka}.
This value is about 2-20 times larger than $l^{\rm avg}_{\rm out}$ obtained in our 3D simulations (see Table~\ref{table:models}).
On the other hand, some theoretical studies have suggested that $l_{\rm out}$ highly depends on the energy conditions of outflowing gas.
The 3D HD simulations by \citet{MacLeod2018aApJ, MacLeod2018bApJ} that investigated gas leakage from the L2 point and subsequent propagation during Roche lobe overflow have shown that 
$l_{\rm out}$ vary from $l_{\rm iso}$ and $l_{\rm L2}$ with changing the equation of state of gas from adiabatic to isothermal.
This indicates that to clarify the nature of angular momentum extraction from binaries,
it is essential to entirely follow the generation of outflows and the subsequent propagation beyond the L2 point, 
self-consistently including gas heating and cooling processes.
This is a crucial subject to be tackled in future studies to understand the long-term orbital evolution of mass-transferring binaries.

\section{Summary}\label{sec:summary}

We have performed multi-dimensional RHD simulations to investigate the mass accretion process onto a stellar-mass BH in a binary undergoing stable mass transfer from the companion star.
The 3D, S2D, and L2D simulations conducted in this paper have assumed different numbers of dimension and covered different radial ranges. 
By combining these three types of simulations, we have unveiled the global accretion dynamics from the L1 point down to about 100 times the Schwarzschild radius of the BH.
In this study, we have especially considered super-Eddington mass transfer with $\mdott \geq 10^3~\medd$, where radiative pressure by diffusive photons in accretion flows can exceed the gravity of the BH, and have evaluated the inward and outward mass fluxes after the system reaches a quasi-steady state.
The key findings of this paper are outlined below:

\begin{itemize}
\setlength{\itemsep}{0.2cm}

\item The inward and outward mass fluxes peak around the outskirts of the accretion disk due to the accumulation of inflowing gas from the L1 point and outflowing gas driven from the inner disk regions.
Inside the disk, the mass fluxes decrease inward due to gradual mass loss by outflows, following $\dot{M} \propto r^{0.6}$.

\item Outflows observed at small radii are primarily associated with convective motions within the accretion disk, thereby being energetically too weak to leave the binary directly. 
However, they undergo a gradual outward transfer
by repeatedly falling back to the disk and undergoing further acceleration by diffusive radiation pressure.
This mechanism propagates energy and momentum outward, finally driving large-scale outflows from the outer disk edge, which eventually escape and extract masses and angular momenta from the binary.

\item Based on the aforementioned findings, we proposed a formula (Eq.~\ref{eq:mdotin}) describing the inflow structure observed in the quasi-steady state.
Our formula predicts a positive correlation between mass transfer rates and mass accretion rates on the central BH, consistent with the results from previous RHD simulations that investigated accretion dynamics down to the ISCO radius.
This agreement suggests that super-Eddington accretion disks generally experience mass loss via radiation-driven outflows starting nearly from the spherization radius and ceasing around a few tens times the Schwarzschild radius.

\item The radiation-driven outflows are heavily anisotropic, affected by the non-axisymmetric external forces within the binary. 
Despite this anisotropy, however, the specific angular momentum of the outflows averaged over all directions, is generally consistent with the value expected in the standard isotropic emission scenario.
The specific angular momentum measured in our simulations serves as a lower limit, 
as we did not account for the additional acceleration of outflows beyond the Roche lobe due to the tidal torque of the binary.

\item While the difference in opacity between solar-abundance and metal-free cases does not significantly impact the nature of radiation-driven outflows, the absence of heavy elements results in a longer viscous timescale, associated with a reduction in diffusive radiation pressure within the disk.

\end{itemize}

These findings provide crucial implications for the orbital decay of binaries during stable mass transfer and, consequently, for the formation of merging binary BHs detected in the LIGO/Virgo/KAGRA collaboration.
We note, however, that the conclusion presented in this paper requires validation in future studies. 
This involves conducting more sophisticated multi-scale simulations that self-consistently track the global structure of outflows and gas circulation and cover scales from the vicinity of the BH to infinity.
Additionally, considering the magneto-hydrodynamic effects, which were not explicitly accounted for in our simulations, would likely modify the nature of outflows.
Moreover, exploring a broader parameter space is an essential extension of this research, helping us to understand the generality or diversity of the properties of mass-transferring binaries.


\section*{Acknowledgements}

We thank K.~Nagamine, S.~Takasao, and T.~Kinugawa for fruitful discussions. 
The numerical simulations were performed with the Cray XC50 at the Center for Computational Astrophysics (CfCA) of the National Astronomical Observatory of Japan.
D.\ T.\ was supported in part by the JSPS Grant-in-Aid for Scientific Research (21K20378 and 22K21349). 
K.\ H.\ was supported by JST FOREST Program (JPMJFR2136) and the JSPS Grant-in-Aid for Scientific Research (20H05639, 20H00158, 23H01169,  23H04900).
K.\ I.\ acknowledges support from the National Natural Science Foundation of China (12073003, 12003003, 11721303, 11991052, and 11950410493) and the China Manned Space Project (CMS-CSST-2021-A04 and CMS-CSST-2021-A06). 
R.\ K.\ acknowledges financial support via the Heisenberg Research Grant funded by the Deutsche Forschungsgemeinschaft (DFG, German Research Foundation) under grant no.~KU 2849/9, project no.~445783058.


\section*{Data availability}

The data underlying this article will be shared on reasonable request to the corresponding author.




\bibliographystyle{mnras}
\bibliography{refs.bib} 



\bsp	
\label{lastpage}
\end{document}